\documentclass[epsf]{emulateapj}
\input epsf

\def\etal{{et al.~}}
\shorttitle{SNRs in Four Irregular Galaxies}
\shortauthors{Chomiuk \& Wilcots}

\begin{document}

\title{A Search for Radio Supernova Remnants in Four Irregular Galaxies}

\author{Laura Chomiuk\altaffilmark{1} \& Eric M. Wilcots\altaffilmark{1}}
\altaffiltext{1}{Department of Astronomy, University of Wisconsin--Madison, 475 N. Charter St., Madison, WI 53706; chomiuk@astro.wisc.edu}

\begin{abstract}
We survey four nearby irregular galaxies for radio supernova remnants (SNRs)  using deep (1$\sigma$ $\sim$\ 20 $\mu$Jy), high resolution ($\sim$20 pc) VLA continuum data at 20, 6, and 3.6 cm. We identify discrete sources in these galaxies and use radio spectral indices and H$\alpha$ images to categorize them as SNRs, H~II regions, or background radio galaxies. Our classifications are generally in good agreement agreement with the literature. We identify a total of 43 SNR candidates: 23 in NGC 1569, 7 in NGC 4214, 5 in NGC 2366, and 8 in NGC 4449.  Only one SNR---the well-studied object J1228+441 in NGC 4449---is more luminous at 20 cm than Cas A. By comparing the total thermal flux density in each galaxy to that localized in H~II regions, we conclude that a significant fraction must be in a diffuse component or in low-luminosity H~II regions. 
\end{abstract}

\section{Introduction}

Supernova remnants (SNRs) are a critical channel through which massive stars feed back on their host galaxies. They provide the kinetic energy for galactic fountains and winds and the heat for the warm interstellar medium (ISM), pollute the ISM with metals, and accelerate cosmic rays. Individual objects have been studied in depth in the Milky Way, but very little is known about SNRs as populations. Surveys of SNRs outside our Galaxy provide an opportunity to understand SNRs in a more systematic way, as SNRs in extragalactic systems are not subject to the high levels of extinction and distance ambiguities present in the Milky Way. 

Other discrete sources like planetary nebulae (e.g., Jacoby 1989, Jacoby \etal 1990), X-ray binaries (Fabbiano 2006), and globular clusters (Brodie \& Strader 2006) can be observed in nearby galaxies and are currently being understood as populations, leading to many advances in fields ranging from the distance ladder to galaxy evolution and cosmology. Our hope is to develop a similar understanding of SNRs, in order to statistically probe the interstellar medium and feedback from massive stars in galaxies with a wide range of masses and Hubble types. 

Surveys for extragalactic SNRs have been carried out for several decades in the radio, optical, X-ray, and infrared (IR). Optical searches generally utilize narrow-band imaging of the [S~II] and H$\alpha$ emission lines, with the expectation that shock-excited SNRs will have elevated [S~II]/H$\alpha$ as compared with photo-ionized H~II regions. This method was pioneered by Mathewson \& Clarke (1973) for use on the Magellanic Clouds, but has since been applied to many nearby galaxies with great success (typical detection rates of 5-100 SNRs per galaxy; e.g., Matonick \& Fesen 1997). Labrie \& Pritchet (2006) recently developed a similar technique using the [Fe~II] and Pa$\beta$ emission lines in the near-infrared (NIR), which should be less vulnerable to dust obscuration than the optical lines. They detect SNRs in three starbursting irregular galaxies, including one which is the subject of this study, NGC 1569. In the X-ray, SNRs can be more difficult to separate from other discrete sources, but several studies have identified SNR candidates based on their soft spectra and lack of time variability (e.g., Pannuti \etal 2000, Pannuti \etal 2002). Martin \etal (2002) surveyed one of our sample galaxies, NGC 1569, in the X-ray and identified an X-ray source as a potential SNR. Finally, nearby galaxies have been surveyed in the radio for SNRs, usually using 20 cm and 6 cm images to constrain discrete sources' spectral indices and separate non-thermal SNRs from H~II regions. Some of the most successful searches of this kind have been in M33 (Gordon \etal 1999) and NGC 6946 (Lacey \etal 1997, Lacey \& Duric 2001). Some of the SNRs in two of the galaxies presented here have been previously studied in the radio (Greve \etal 2002 in NGC 1569 and Vukoti\'{c} \etal 2005 in NGC 4214; see Section 2). In most cases, the different wavelength techniques detect distinct kinds of SNRs (see Pannuti \etal 2002 and Duric 2000), and galaxies should be surveyed at as many wavelengths as possible in order to achieve a complete census of SNRs.
 
Our goal is to study the radio luminosity function of SNRs, and therefore in this study, we focus on SNRs selected by their 20 cm emission. We present a catalog of the discrete radio sources in four irregular galaxies-- NGC 1569, NGC 2366, NGC 4214, and NGC 4449. Radio-bright SNRs typically have diameters of $\sim$10-100 pc (Gordon \etal 1999), corresponding to 0.7-7$^{\prime\prime}$ at a distance of 3 Mpc. Therefore, for nearby galaxies, they will appear as discrete (point-like or slightly resolved) sources when observed with most ground-based observatories. The radio continuum of star-forming galaxies is composed of three components: thermal free-free emission arising from H~II regions, synchrotron emission from supernova remnants (SNRs), and diffuse non-thermal emission. The high-resolution observations we present here mostly resolve out the latter diffuse component, but it remains an important task to separate H~II regions from SNRs. 

This work makes progress on our understanding of four irregular galaxies by bringing together data from the VLA taken over the past 25 years and achieving sensitivites of $\sim$20 $\mu$Jy. We select discrete sources based on their 20 cm emission and measure their radio spectra using 20 cm, 6 cm, and 3.6 cm flux densities. We characterize them as SNRs, H~II regions, or background galaxies, and then discuss some individual sources of note and compare our sample with sources in the literature. Finally, we consider future applications for samples like this one. This study seeks to increase the number of known radio SNRs, with the hope that we can ultimately extract the statistical properties of SNR populations in galaxies.
\section{Survey Overview}

In this publication, we present data on the four galaxies in Table 1. These galaxies were chosen to be nearby ($<$ 5 Mpc) and to display irregular morphologies. Their B-band absolute magnitudes span 2 magnitudes, sampling the upper end of dwarf galaxy masses. These galaxies are ideal sources for this study because each has been observed at high-resolution (1-4$^{\prime\prime}$) at three wavelengths: 3.6, 6, and 20 cm. All radio data used here were obtained with the Very Large Array (VLA) of the National Radio Astronomy Observatory (NRAO)\footnote{The National Radio Astronomy Observatory is a facility of the National Science Foundation operated under cooperative agreement by Associated Universities, Inc.}. This study imposes a strict selection criterion, identifying all discrete radio sources with flux measurements $>$3$\sigma$ at 20 cm, and then systematically characterizes them using the ratios between 20 cm,  6 cm, and 3.6 cm flux densities. 

Greve \etal (2002) surveyed NGC 1569 with MERLIN at very high resolution ($\sim$0.2$^{\prime\prime}$) at 20 cm, and supplement their data with 6 cm VLA images in order to calculate spectral indices. Their high resolution images are predominantly sensitive to sources with very high surface brightness, tending to be the youngest, most compact SNRs. They also include several sources which are detected on VLA 20 cm images, but unfortunately the VLA data was never formally published so we can not know how these sources were chosen. Therefore, our data complements the work of Greve \etal (2002) by systematically detecting more extended and/or less luminous SNRs. Vukoti\'{c} \etal (2005) search for SNRs in NGC 4214 using VLA data, but they do not state their detection criteria and therefore their sample does not have a clear and reproducible selection function. They detect three SNR candidates which are discussed further in Section 6.

At the VLA, the primary beam width at half power (equivalent to the field of view) is $\sim$30$^{\prime}$ at 20 cm, 9$^{\prime}$ at 6 cm, and 5.3$^{\prime}$ at 3.6 cm. Because our scientific goals require spectral index information for our sources, a flux density measurement must exist at at least two frequencies. Therefore, our survey area for each galaxy should cover a circular area with diameter of 9$^{\prime}$. However, due to bandwidth smearing (see Section 3), the 20 cm image often ends up limiting the survey area. 

The optical angular diameters of the four galaxies are listed in Table 1, as defined by the B = 25 mag arcsec$^{-2}$ isophote (R$_{25}$; Nilson 1973). NGC 1569, NGC 2366, and NGC 4449 are well-covered to approximately one optical radius or more by the data presented here, and therefore our survey should include most of the radio sources in these galaxies. Unfortunately, this is less true for NGC 4214; it has the largest optical diameter and is severely affected by bandwidth smearing.

Integrated 20 cm flux densities for the four galaxies are also listed in Table 1, taken from Condon (1987). These flux densities were obtained with the VLA in its D configuration. Given the angular sizes of our galaxies, all of the galaxies' flux should be included in these measurements. Finally, the H$\alpha$ integrated luminosities of the galaxies are also provided, taken from Hunter \& Elmegreen (2004) and Hunter \etal (1999) and adjusted for the distances in Table 1. For low-luminosity galaxies like those studied here, L$_{H\alpha}$ is a good proxy for star formation rate (Hunter \& Elmegreen 2004).

\section{Data Reduction}

All VLA data are publicly available through the NRAO Archive, obtained by other investigators between the years of 1983 and 2002, with the exception of the C-band data for NGC 2366 which were obtained by the present authors in 2006. The details of the data for each galaxy, and the images created from this data, are listed in Tables 2 and 3. 

All data were observed using 2 IFs and 2 polarizations. Most data sets were observed in ``continuum mode" with one channel of 50 MHz width per IF and per polarization yielding an effective bandwidth of 200 MHz. However, there are a few exceptions. Both the 6 and 20 cm data for NGC 4449 were taken in 7 channel format, with 3.125 MHz per channel, for an effective bandwidth of 87.5 MHz. The C-band data for NGC 2366 were observed using a 3-channel correlator configuration, with 12.5 MHz bandwidth per channel, for a total bandwidth of 150 MHz. 

Although the continuum correlator configuration maximizes bandwidth, the images suffer badly from chromatic aberration (bandwidth smearing) away from the phase center, compromising spatial resolution by smearing out all sources in a radial direction and decreasing sensitivity by diluting the source signal over more sky noise. Bandwidth smearing is most severe at higher resolution; sources which are 80$^{\prime\prime}$ from the center of an A-configuration 20 cm image will be widened by a factor of two and suffer a 50\% loss in peak brightness. Luckily, NGC 1569 has a small angular size (see Table 1), so we are able to survey the entire optical disk at high sensitivity. This is not true for NGC 4214, which has a large optical diameter of $\sim$10$^{\prime}$; we are only sensitive to its central 3$^{\prime}$. In a B-configuration image, sources will not be equivalently smeared until a radius of 3$^{\prime}$ or so. NGC 2366 was observed in B-configuration, and like NGC 1569, its angular size is well matched to the radius where bandwidth smearing becomes a severe issue. NGC 4449 was observed in multi-channel mode and therefore is not affected by chromatic aberration, allowing one to image the entire 30$^{\prime}$ L-band primary beam. Bandwidth smearing is not a concern at C- or X-band.

The data were flagged, calibrated, and imaged in the usual way using classic AIPS. For a given galaxy and receiver, the data from multiple observing programs and/or configurations were concatenated in the \emph{uv}-plane. During the imaging step, we tweaked the \emph{robust} parameter so that the L-, C-, and X-band images for each galaxy would have similar resolutions. We also corrected for primary beam attenuation. Finally, we matched the resolution more precisely by smoothing the images to the values listed in Table 3. The spatial resolutions (in pc) of these images are also listed assuming galaxy distances as in Table 1. These resolutions are all comparable to the diameters of radio-bright SNRs in M33 (Gordon \etal 1999). The sensitivities of the images after smoothing can also be found in Table 3. All further analysis is performed on these smoothed matched-resolution images, which are shown in Figures 1, 3, 5, and 7.

\section{Data Analysis}

Discrete sources were detected in the 20 cm images using SExtractor (Bertin \& Arnouts 1996), a flexible code which can detect both point and extended sources. We requested it to detect all regions with five adjacent pixels brighter than 4 times the image noise. We then visually checked the source list, removing a few clear cases of spurious detections and occasionally adding sources that were missed by the algorithm.

Flux densities were measured for the radio sources by fitting Gaussian functions to them using the AIPS task \emph{jmfit}. There are several features of \emph{jmfit} which make it the best candidate for the measurements:
\begin{enumerate}
\item Unlike aperture photometry, Gaussian fitting does not impose a certain size for the object. Many of our sources are unresolved, but then again, many are extended and still others are stretched out by bandwidth smearing. It is impossible to calculate aperture corrections for such a varied sample. The alternative to using an aperture correction is to measure the flux density in a large aperture ($>>$ FWHM), but many of our sources have very low signal-to-noise (S/N) and will be lost in the sky noise if measured in such a large aperture.
\item Gaussian fitting can fit multiple overlapping sources simultaneously, a feature which is very useful in crowded regions like the disk of NGC 1569. Again, this is impossible to do well with aperture photometry unless one uses a small aperture and a well-defined aperture correction (which is impossible to determine for our data; see above).
\item It can account for bandwidth smearing.
\end{enumerate}
Of course, the many free parameters involved in Gaussian fitting can also add uncertainty to our analysis. For sources with low S/N or in regions which are crowded, complex, or bright with diffuse emission, \emph{jmfit} may not fit the intended emission and/or it may make an outlandish guess at the width of the Gaussian. In such cases, it is necessary to fix some parameters (the Gaussian position or width) in order to simplify the fit. \emph{Jmfit} also allows one to fit the background around the Gaussian. In most cases where sources are fairly isolated, the background was simply characterized as a constant. For some sources in crowded bright regions, the background was described as a gradient or a curve+slope.

For a given source of interest, we carefully chose a surrounding section of the 20 cm image that was large enough for the background to be characterized accurately, but small enough that this characterization could be fit by a simple function. We then fit a Gaussian to the source (or a few Gaussians to multiple overlapping sources) in the 20 cm image. Next, we used the same image section in the 6 and 3.6 cm images and attempted to fit corresponding sources at these higher frequencies. If the location or width of the 6 or 3.6 cm Gaussian was significantly different than the 20 cm fit (implying low S/N or confusion), then the location and/or width were held fixed at the values determined from the 20 cm image. Residual images were inspected to ensure that the Gaussian fitting was reasonable. If there was clearly no source present above the noise in the 6 or 3.6 cm image, then the location and width of the Gaussian were set to those measured at 20 cm and \emph{jmfit} was used to measure the local noise.

A source makes it into our sample if its 20 cm flux density is $>$3$\sigma$ (where sigma is the error in the integrated flux density given by the Gaussian fit). These sources are marked with circles in Figures 1-8, and their positions and measured flux densities are presented in Tables 4 through 7. Positions are accurate to approximately $\sim$0.5$^{\prime\prime}$. Due to the smaller primary beam at X-band, some sources do not have measured 3.6 cm flux densities. Additionally, if a source is measured at $<$3$\sigma$ significance at 6 or 3.6 cm, an upper limit to its flux density is given in the tables. The upper limit corresponds to 3$\sqrt{2}$ $\sigma$. The major and minor axes for the sources are also presented in these tables in arcseconds. These include bandwidth smearing and therefore may not reflect the true angular sizes of all sources.

Spectral indices (defined as S $\propto \nu^{\alpha}$) are calculated for each source and are listed in Tables 4-7. If a source is measured at all three frequencies, its spectral index is found by minimizing the $\chi^{2}$ statistic in a linear fit to log $\nu$- log S, and the error in the spectral index is the error in the line's slope. If flux densities at only two frequencies are measured, then the spectral index is found by taking the ratio of these two measurements:
$$ \alpha_{1-2} = {{log (S_{1} / S_{2})} \over {log (\nu_{1} / \nu_{2})}}  $$
where $S_{1}$ and $S_{2}$ are flux densities measured at two different frequencies, and $\nu_{1}$ and $\nu_{2}$ are these frequencies (as listed in Table 3). In this case, errors in $\alpha$ are simply found by propagating the photometric errors. Finally, if the only formal measurement is at 20 cm, with upper limits on the flux density at 6 cm and 3 cm, we use the data to place an upper limit on the spectral index. In this case, the spectral indices recorded in Tables 4-7 represent the censored value between 20 cm and either 6 cm or 3.6 cm, whichever places the most stringent limits on the spectral index.

\section{Source Classification}

There are three types of sources which are thought to contribute significantly to the population of discrete radio sources in galaxies: background radio galaxies, H~II regions, and SNRs. H~II regions should display flat radio spectra characteristic of bremsstrahlung ($\alpha \sim$ -0.1), whereas SNRs radiate in non-thermal synchrotron emission, and have typical radio spectral indices of $\alpha \sim$ -0.45 (Condon 1992). Background radio galaxies are also powered by synchrotron emission, and should host relatively steep spectra. It is important to keep in mind, however, that the radio spectra of both SNRs and radio galaxies have a wide range of shapes and can be quite flat, masquerading as H~II regions (Gordon \etal 1999). 

Currently, the best way to separate out these classes of objects is a combination of radio spectral indices and optical imaging of the H$\alpha$ emission line (Pannuti \etal 2000). Spectral indices can be used to separate out thermal spectra from synchrotron spectra. We impose the criterion that a source with $\alpha >$ -0.2 is thermal, and $\alpha \leq$ -0.2 is non-thermal (Pannuti \etal 2002). Spectral indices are better constrained here than in many past searches for SNRs (Lacey \etal 1997, Gordon \etal 1999, Pannuti \etal 2000, 2002), because for many sources we have measured flux densities at three frequencies instead of just two.

H$\alpha$ imaging is then used to separate background sources from sources in the target galaxy, because H$\alpha$ filters are generally quite narrow ($<100$\AA) and background radio galaxies tend to be at significant redshifts. Any H$\alpha$ emission that a background galaxy emits will be redshifted out of the narrow-band filter centered on z $\sim$ 0. In NGC 4214, NGC 2366, and NGC 4449, we look for clear discrete enhancements in the H$\alpha$ emission around the vicinity of the radio source. However, in NGC 1569, we use a slightly different criterion, because the H$\alpha$ emission is very high-surface brightness all throughout the galaxy's star-forming disk and would easily overwhelm discrete H$\alpha$ sources. Therefore, if radio sources are within this high-surface brightness disk (major axis $\sim$ 1.5$^\prime$), we consider them constituents of NGC 1569. This is a reasonable assertion because, given our sensitivity and the number counts from Condon (2007), we would expect less than one background source within the bright disk.

An unfortunate complication is that many radio SNRs do not emit brightly in H$\alpha$ (Pannuti \etal 2000), and these sources may be confused with background galaxies. However, it appears that we are not mistakenly identifying many SNRs as background sources. Our survey is sensitive to sources with flux densities down to $\sim$0.10 mJy, and Condon (2007) implies that there should be on average 0.26 background sources per square arcminute brighter than this flux limit. In NGC 1569, 1.3 background sources are predicted within R$_{25}$; we find two. In NGC 2366, 6.4 background sources are predicted and we find four, while in NGC 4449, one would expect 5.5 background sources and four are detected. Therefore our results are in good agreement with the Condon (2007) number counts, implying that our methods neither dramatically over- or under-estimate the number of background sources. 

We obtained H$\alpha$ narrow-band images for NGC 1569, NGC 4214, and NGC 2366 from Hunter \& Elmegreen (2004), respectively pictured in Figures 2, 4, and 6. The H$\alpha$ image for NGC 4449 is from Hunter \etal (1999) and can be seen in Figure 8. A world coordinate system was appended to these images using Koords (part of the Karma\footnote{http://www.atnf.csiro.au/computing/software/karma/} software package; Gooch 1996) and by comparing them with images from the Digitized Sky Survey\footnote{The Digitized Sky Survey was produced at the Space Telescope Science Institute under U.S. Government grant NAG W-2166. The images of these surveys are based on photographic data obtained using the Oschin Schmidt Telescope on Palomar Mountain and the UK Schmidt Telescope. The plates were processed into the present compressed digital form with the permission of these institutions.}. We then searched for H$\alpha$ counterparts to our radio sources. Sources with H$\alpha$ counterparts are marked with red, magenta, or cyan circles in Figures 1-8 and marked with a ``Yes" in Tables 4-7. If there is no H$\alpha$ emission around a source, it is marked with a green circle in the figures and labeled ``No" in the tables. If a source is marked with ``---", then it falls outside the H$\alpha$ imaging. 

In addition, we checked the sources which fall outside R$_{25}$. It is highly improbable to find an H~II region or SNR outside the optical diameter, and indeed, all sources found at these large radii are consistent with background radio galaxies. In some cases, the H$\alpha$ image did not extend to these large radii, and so we could not perform the usual check described above. However, we still considered such objects background sources as long as they were beyond R$_{25}$.

Finally, Tables 4-7 contain our source classifications: Bkg, H~II, or SNR. It is important to bear in mind that these are only guesses from the current limited information, and more certain classification would require additional data like optical/near-IR spectroscopy (see Section 6, where for example, [Fe~II] and Pa$\beta$ narrow-band imaging of NGC 1569 by Labrie \& Pritchet (2006) help us confirm SNR candidates). Below is a summary of our classification criteria: 
\begin{itemize}
\item Bkg: has no apparent H$\alpha$ emission associated with it. If the source is located beyond the galaxy's optical diameter, it is marked with ``$>$R$_{25}$".
\item SNR: has $\alpha \leq$ -0.2 and H$\alpha$ emission associated with it.
\item H~II: has $\alpha >$ -0.2 and H$\alpha$ emission associated with it.
\item SNR/H~II: has H$\alpha$ associated with it and an upper limit on its spectral index that is consistent with either an H~II region or SNR.
\end{itemize}

Also, it is important to bear in mind that for each galaxy, the L, C, and X band data were collected at different epochs. Several sources (like N1569-02) have spectra which are difficult to explain physically, and may result from variable flux densities. In addition, the luminous well-studied SNR in NGC 4449 (N4449-12) was shown to be fading in the radio by Lacey \etal (2007). However, this object is unusually young ($\sim$100 years; Blair \etal 1983) amongst radio SNRs with typical lifetimes of $\sim$10$^{4}$ years (Duric \etal 1993). The radio emission of most SNRs does not fade significantly in a few decades time, and therefore this data set is appropriate for all but the youngest, most compact SNRs. Of the three classes of objects, only background radio galaxies are expected to vary significantly in brightness.
\section{Individual Sources of Note}

Many objects in our sample have been discussed in the literature or deserve a few more words. In particular, Labrie \& Pritchet (2006) identified 11 SNR candidates in NGC 1569 from narrow-band NIR imaging of both the [Fe~II] and Pa$\beta$ lines. [Fe~II] emission is a good indicator of shocks, and the line ratio [Fe~II]/Pa$\beta$ will be higher in SNRs than in H~II regions (Mouri \etal 2000). Six of their candidates overlap with sources in our sample, and we independently categorize four (N1569-10) as SNRs. We also compare our sources with the X-ray survey of NGC 1569 by Martin \etal (2002).

\emph{N1569-08.---} Greve \etal (2002) detected and measured this source with the VLA (labeling it VLA-11) but not with higher-resolution ($\sim$0.2$^{\prime\prime}$) MERLIN data, and therefore the authors claim that it is extended. They measure a non-thermal radio spectrum, and hypothesize that it is a low-surface-brightness SNR. 

\emph{N1569-10.---} This source corresponds to a [Fe~II] source and SNR candidate S010 in Labrie \& Pritchet (2006). We categorize it as an H~II region, but a glance at Table 4 shows that its spectral index is right at the edge of our cut-off between thermal and non-thermal spectra. Given the near-IR imaging results, we believe this source is probably an SNR with a relatively flat spectrum.

\emph{N1569-11.---} This source is a MERLIN detection (M2; Greve \etal 2002) with a small angular size and non-thermal spectrum. It is hypothesized to be a radio supernova (RSN) or compact SNR, in agreement with our results.

\emph{N1569-13.---} This source is also a MERLIN detection (M1; Greve \etal 2002) and corresponds to the H~II region Waller No. 2 (Waller 1991). We agree that it displays a thermal spectrum.

\emph{N1569-14.---} This is a compact MERLIN source (M3; Greve \etal 2002), probably a RSN or small SNR. It was also identified as an SNR candidate in the NIR (S009; Labrie \& Pritchet 2006), and is therefore almost surely an SNR. It is also positionally coincident with X-ray source CXOU 043046.9+645106 in Martin \etal (2002), which is characterized as an X-ray binary based on low S/N comparisons of soft, medium, and hard X-ray fluxes. Whether the X-ray source is actually associated with this object, or if it is just a chance alignment, is unclear. It is extremely unlikely that the radio source is an X-ray binary; the radio luminosities of such sources are far too low to be observed at extragalactic distances (Fender \& Hendry 2000).

\emph{N1569-15.---} This is a VLA source (VLA-16) in Greve \etal (2002), who measure it to have a non-thermal spectrum, in contradiction with the thermal spectrum we measure for this object. Even in our relatively low-resolution VLA images, this source has a diffuse, extended morphology. Greve \etal (2002) measure its flux density by assuming it is a point source, thereby dramatically underestimating it. There may be a compact element emitting non-thermally which they measure, but the bulk of the flux coming from N1569-15 is thermal.
 
\emph{N1569-16.---} This source (here categorized as an SNR) is coincident with X-ray source CXOU 043047.5+64505, which is most likely an X-ray binary according to Martin \etal (2002).
 
\emph{N1569-19.---} In Greve \etal (2002), this source was detected with the VLA but not MERLIN (VLA-8). It was also measured to have a non-thermal spectrum, and was therefore characterized as an extended SNR. The source is also a NIR SNR candidate (S007; Labrie \& Pritchet 2006). We measure this source to have a spectral index $\alpha$ = -0.19, right at the brink between thermal and non-thermal sources. We are forced to classify it as an H~II region in Table 4, but previous work solidly implies that it is an SNR.

\emph{N1569-24.---} This is the MERLIN detection M4 (Greve \etal 2002) and the H~II Region Waller No. 5 (Waller 1991). 

\emph{N1569-27.---} This source was detected by MERLIN (M5) and described as a mixture of thermal and non-thermal emission by Greve \etal (2002). We measure a non-thermal spectrum for it, and it is listed as an SNR candidate (S008) by Labrie \& Pritchet (2005), implying that it is indeed an SNR.

\emph{N1569-28.---} This source is NIR SNR candidate S002 (Labrie \& Pritchet 2006).

\emph{N1569-36.---} X-ray source CXOU 043053.3+645044 is coincident with this source and is characterized as an X-ray binary by Martin \etal (2002).

\emph{N1569-38.---} This source is a spectacular example of an extragalactic SNR. It is a non-thermal VLA/MERLIN source (M6; Greve \etal 2002), a NIR SNR candidate (S001; Labrie \& Pritchet 2006), and an X-ray source (CXOU 043046.9+645106; Martin \etal 2002). Martin \etal also point out that it has a shell-like morphology in [S~II] and [O~III] narrow-band images.

\emph{N1569-42.---} This source is detected in the X-ray (CXOU 043058.0+644910) and categorized as an active galactic nucleus (AGN) by Martin \etal (2002), consistent with our "Bkg" description.

\emph{N1569-45.---} This source corresponds to X-ray source CXOU 043113.2+645229 and is characterized as a star based upon its ratio of X-ray to optical flux by Martin \etal (2002). Because of its bright radio flux density, we assert that it is more likely an AGN.

\emph{N4214-08 \& N4214-10.---} These two sources fall in the region which was identified as an SNR by MacKenty \etal (2000), based upon a non-thermal radio spectral index, optical emission line ratios, and kinematics. The exact location of the SNR is unclear due to the poorer resolution of the radio data in MacKenty \etal (they used an aperture 4$^{\prime\prime}$ in radius).

\emph{N4214-14, N4214-16, \& N4214-19.---} These three sources correspond to SNR candidates $\alpha$, $\beta$, and $\rho$ in Vukoti\'{c} \etal (2005). We disagree with this work's claims that N4214-14 and N4214-16 have non-thermal spectra, and categorize them as H~II regions. These discrepancies are difficult to explain, but they may be due to the fact that Vukoti\'{c} \etal used a different technique for measuring 20 cm flux densities than they used for measuring the 6 and 3.6 cm flux densities. We agree with Vukoti\'{c} \etal that N4214-19 has a non-thermal spectrum and is a likely SNR.

\emph{N2366-10.---} This source is part of the bright H~II region complex NGC 2363 (Kennicutt \etal 1980), and is labeled NGC 2363 A by Yang \etal (1994). We disagree with the assertion of Yang \etal that it displays a non-thermal radio spectrum and is therefore dominated by radio SNRs. Our data, which include three frequencies and are significantly more sensitive, imply a spectral index of $\alpha$ = -0.13 (easily explained by simple bremsstrahlung). However, this does not necessarily imply that there are no SNRs lurking in this bright H~II region. Roy \etal (1991) measured double-peaked [O~III] profiles near this source and Yang \etal detected large H$\alpha$ velocity widths in this region. 

\emph{N2366-11.---} This source is also part of the H~II region NGC 2363, and is called NGC 2363 B by Yang \etal (1994). They measure a non-thermal spectrum for this source, too, and again we disagree with them by measuring a thermal spectrum. It is worth pointing out that there is significant spatial overlap between N2366-10 and N2366-11 at our resolution (and at the resolution of Yang \etal), and therefore the spectral indices of these two objects are not completely independent.

\emph{N2366-14.---} This object has a clear double-lobed jet morphology, and is therefore almost certainly a background radio galaxy.  

\emph{N4449-06, N4449-07, N4449-09, N4449-10, N4449-16, \& N4449-18.---} Six of our sources in NGC 4449 are identified as obscured massive young stellar clusters by Reines \etal (2008), and correspond to their sources \#3, 5, 7, 11, 18, and 23 respectively. For all sources but N4449-07, we agree with Reines \etal that they exhibit thermal spectra. However, Reines \etal identified N4449-07 as ``likely thermal" based on an X-band flux density and upper limits at 6 cm and 1.3 cm. We identify this source as non-thermal based on flux density measurements at 20, 6, and 3.6 cm. 

\emph{N4449-12.---} This bright object, otherwise known as J1228+441, is one of the most famous extragalactic SNRs. It was discovered by Seaquist \& Bignell (1978) and has been studied extensively in the X-ray (Summers \etal 2003, Patnaude \& Fesen 2003, Vogler \& Pietsch 1997),  optical (Milisavljevic \& Fesen 2008 and references therein), ultraviolet (Blair \etal 1984), and radio (Lacey \etal 2007). This object is a young ($\sim$100 years old) oxygen-rich remnant (Blair \etal 1983). Unfortunately, our spectral index measurement is not reliable for this source because the data was taken over 18 years during which the radio emission was still fading (Lacey \etal 2007).

\section{Supernova Remnants}

We find a total of 43 radio SNR candidates: 23 in NGC 1569 (plus two which may be SNRs or H~II regions), 7 SNR candidates (plus 3 SNR/H~II) in NGC 4214, 5 in NGC 2366, and 8 in NGC 4449. These sources are shown in more detail in Figure Set 9, with 20 cm contours overlain on H$\alpha$ grey-scale as can be seen in Fig. 9. We now compare our SNR luminosities with the model of Lisenfeld \& V\"{o}lk (2000), who predict that 10\% of a galaxy's synchrotron emission should be in SNRs. 

As can be seen in Table 8, the SNRs detected in NGC 1569 comprise 17.7$\pm$0.5 mJy (or 18.4$\pm$0.6 mJy if one includes sources categorized as SNR/H~II), or 4\% of the total 20 cm flux density. In NGC 4449, SNRs sum to 12.5$\pm$0.2 mJy-- 5\% of the galaxy's total 20 cm flux density, and in NGC 2366, SNRs sum to 1.7$\pm$0.1 mJy (7\% of total). Unfortunately, as we do not survey the entire optical extent of NGC 4214, we can not perform a similar calculation for it.

Based on integrated radio flux densities of NGC 1569 ranging from 37 MHz to 24.5 GHz, Lisenfeld \etal (2004) found a thermal fraction of $\sim$0.23 at 1.49 GHz. This means that 6\% of NGC 1569's synchrotron emission is in discrete sources. In NGC 4449, Niklas \etal (1997) used data at six different frequencies and found a thermal fraction of 0.23$\pm$0.03 at 1.49 GHz, implying that SNRs contribute 6\% of this galaxy's non-thermal radio flux density. Unfortunately, no thermal fraction exists in the literature for NGC 2366.

The reader must bear in mind that the SNR fractions found here are lower limits. Due to our sensitivity limits, our survey does not detect the faintest SNRs, and we can not know how significantly low-luminosity SNRs contribute to the total flux of SNRs in a galaxy, because the LF of radio SNRs is currently unknown. In a future paper, we will attempt to measure the SNR LF and, integrating under it, constrain the total 20 cm luminosity of the SNR population in galaxies. In addition to faint SNRs, we will miss plerionic remnants with flat radio spectra (Weiler \& Sramek 1988), aged SNRs with very low surface brightness, and radio SNRs with no H$\alpha$ counterpart (because they are not located near star-forming regions or do not host optical emission lines themselves). However, our current data is consistent with the prediction of Lisenfeld \& V\"{o}lk (2000).

Cas A, the most luminous SNR in the Milky Way, has a 20 cm flux density of 2720 Jy (Green 1998) and a distance of 3.4 kpc (Reed \etal 1995). At the distance of NGC 1569, Cas A would have a flux density of 8.3 mJy. The brightest SNR we measure in NGC 1569, N1569-38, is therefore only half as luminous as Cas A. Despite the recent high levels of star formation in NGC 1569, it contains no SNRs which rival the luminosity of Cas A.

At the distance of NGC 4214, Cas A would have a flux density of 3.6 mJy. The brightest SNR candidate we measure in this galaxy is 1.4 mJy, or 40\% the luminosity of Cas A. One should recall, however, that we did not survey the entire extent of NGC 4214. Similarly, the brightest SNR candidate in NGC 2366 (N2366-12) has a flux density of 0.9 mJy, or 30\% the luminosity of Cas A. 

The bright well-studied SNR in NGC 4449 (N4449-12) is extraordinarily luminous-- five times more luminous than Cas A. The second most luminous SNR in this galaxy is N4449-14, with a luminosity 80\% that of Cas A. It is remarkable that NGC 4449 has two SNRs which are both significantly more luminous than any of the SNRs in NGC 1569 or NGC 2366. It is possible that this is simply due to its higher star formation rate (see H$\alpha$ luminosities in Table 1), a suggestion put forward by Lacey \& Duric (2001) to explain why the most luminous SNRs in NGC 6946 are $\sim$10 times brighter than those in M33. A more detailed study of the causes for this phenomenon will be the subject of a future paper.

\section{Thermal Radio Emission}

Even with the deep radio continuum images presented here, we are only sensitive to the very brightest H~II regions. Except in the case of heavily obscured galaxies, H$\alpha$ surveys are much more sensitive to H~II regions. Assuming a nebular temperature of 10$^{4}$ K and using the equations detailed in Caplan \& Deharveng (1986) to convert between H$\alpha$ flux and radio free-free flux density, we note that a faint 20 cm source with a flux density of 0.1 mJy is expected to emit an H$\alpha$ flux of $8\times10^{-14}$ ergs cm$^{-2}$ s$^{-1}$. In contrast, most H$\alpha$ surveys can detect H~II regions with fluxes significantly below $10^{-14}$ ergs cm$^{-2}$ s$^{-1}$ (e.g., Kennicutt \etal 1989, Youngblood \& Hunter 1999).

Therefore, it is not surprising that at 20 cm we only identify a handful of H~II regions in each galaxy, while the H$\alpha$ suvey of Kennicutt \etal (1989) detects 97 H~II regions in NGC 2366 and 136 H~II regions in NGC 4449. Using \emph{Hubble Space Telescope} narrow-band photometry, Buckalew \& Kobulnicky (2006) identified an impressive 1018 H~II regions in NGC 1569. When we transform the H$\alpha$ fluxes to 20 cm thermal flux densities using the conversion described in Caplan \& Deharveng (1986), we find good agreement between the flux densities of the brightest H$\alpha$-selected regions and the radio H~II regions. In the cases of NGC 2366 and NGC 4449, which Kennicutt \etal (1989) did not correct for internal extinction, this implies that dust extinction is not severely attenuating the H$\alpha$ emission. In the case of NGC 1569, whose H~II regions were dereddened using C(H$\beta$), this implies that the Balmer decrement correction serves well.

Table 8 lists the total 20 cm flux density in H~II regions detected in this work and the total 20 cm flux density of H$\alpha$-selected H~II regions (again, assuming the above-described conversion factor between H$\alpha$ emission and bremsstrahlung). Both of these factors are lower limits because neither survey mode is sensitive to all H~II regions in a given galaxy. It is difficult to know how much flux we are missing because the low-luminosity end of the H~II region LF is poorly known. Buckalew \& Kobulnicky fit a power-law to the H~II region LF in NGC 1569, down to their approximate completeness limit of $log\ L_{H \alpha}$ = 37.22. If we extrapolate this power law to lower luminosity H~II regions, integrating down to $log\ L_{H \alpha}$ = 36, the total luminosity in H~II regions more than doubles. Such an estimate might be too high, however, if the LF turns over as Youngblood \& Hunter (1999) suggest it may in many irregular galaxies. Clearly, the shape of the LF at low luminosities has a powerful impact on the total luminosity of H~II regions. Yet another reason that the summed fluxes of the H$\alpha$-selected H~II regions are lower-limits is that they have not been corrected for extinction (or in the case of NGC 1569, C(H$\beta$) is known to underestimate extinction; e.g. Melnick 1979).

Although by count our 20 cm survey is missing the vast majority of H~II regions, we appear to be missing very little of the total flux density in NGC 2366 H~II regions, and approximately half of the H~II region flux density in NGC 1569 and NGC 4449. This simply shows that a large fraction of a galaxy's H~II-region energy can be found in a small handful of the brightest regions.

We can also investigate what fraction of the total thermal emission is in detected H~II regions. Using the Lisenfeld \etal (2004) thermal fraction for NGC 1569, a thermal flux density of 94.5 mJy is implied (assuming the integrated 1.49 GHz flux density of Condon 1987, listed here in Table 1). The radio-selected H~II regions contribute at most 24.7$\pm$0.6 mJy (if we also include sources with SNR/H~II designation), or $\sim$6\% of the integrated 1.49 GHz flux density. If we instead use the H$\alpha$-selected sample from Buckalew \& Kobulnicky (2006) and the conversion to radio flux density in Caplan \& Deharveng (1986), the H~II regions then contribute 13\% of the total 20 cm flux density. This implies that 43\% of NGC 1569's thermal flux is in diffuse form or in faint H~II regions.

In NGC 4449, the total flux density of H~II regions detected here is 6.0$\pm$0.2 mJy, or only 2\% of the total 20 cm flux density from Condon (1987). The H$\alpha$-selected H~II region sample of Kennicutt \etal (1989) sums to 13.7 mJy or 5\% of the total flux density. Therefore, in NGC 4449, assuming the Niklas \etal (1997) thermal fraction, 78\% of the thermal flux density remains unaccounted for. 

Youngblood \& Hunter (1999) performed similar calculations on their sample of irregular and blue compact dwarf galaxies, comparing the H$\alpha$ flux in discrete H~II regions with the integrated H$\alpha$ flux of each galaxy. They calculate that 49\% of NGC 1569's thermal flux is in H~II regions, in rather good agreement with our result. NGC 4449 is not in their sample, but their results imply that a galaxy with 22\% of its H$\alpha$ emission measured to be in H~II regions is on the low side, but is not particularly unusual.

Clearly, our 20 cm observations are missing significant amounts of thermal flux. Some of this flux is in lower-luminosity H~II regions, but we can not rule out that there is also a diffuse component to the thermal radio flux. Such a component would not be surprising, as H$\alpha$ imaging of galaxies often shows diffuse emission, and free-free radio emission should scale with H$\alpha$. 

\section{Conclusion}

In four nearby irregular galaxies, we have carried out systematic searches for radio SNRs. We detect discrete radio sources at 20 cm on deep high-resolution VLA images, and then use high-resolution VLA data at three radio frequencies to constrain the sources' radio spectra, and H$\alpha$ narrow-band imaging to imply membership in the target galaxy. Our survey uses well-defined criteria to select and measure sources, and it is also an improvement on many radio SNR surveys because it uses three radio frequencies, providing additional constraint on sources' spectral indices.

We measure a total of 43 radio SNR candidates: 23 in NGC 1569, 7 SNR candidates in NGC 4214, 5 in NGC 2366, and 8 in NGC 4449. Only one of these-- the well-studied SNR in NGC 4449-- is more luminous than Cas A. A prominent SNR in NGC 1569 (here labeled N1569-38) has been confirmed in the radio, IR, optical, and X-ray and deserves detailed study. The summed flux densities of our candidate radio SNRs are $\sim$5\% of the integrated 20 cm flux densities for our sample galaxies. This is a lower limit on the contribution of SNRs, as we are not detecting faint SNRs which may be great in number. 

By comparing our summed flux densities of H~II regions with what is expected from the integrated radio spectra of our sample galaxies, we know that we are missing significant amounts of thermal flux. Much of this flux is in faint H~II regions, but we also can not rule out that some fraction of the thermal emission is arising from a diffuse warm component of the ISM. Multi-configuration images with full \emph{uv} coverage are really needed to study the relative contributions and relationships between H~II regions, discrete radio SNRs, the diffuse warm ISM, and the diffuse relativistic ISM. 

The EVLA will dramatically improve surveys like this one through better sensitivity, higher frequency resolution (so that bandwidth smearing is not an issue), and broader frequency coverage (providing more information on a source's radio spectrum). With the EVLA, deep surveys of a wide range of nearby galaxies will allow us to consider SNRs as populations and probes of the ISM in their parent galaxy. One of the first steps in understanding radio SNR populations is measuring the SNR LF. A future paper will summarize past VLA results and attempt to ascertain if the SNR LF is invariant, and if it varies, what the variations depend upon. Additionally, in the future we will be able to accurately constrain the relative contributions of radio SNRs, H~II regions, and diffuse synchrotron to a galaxy's total radio continuum. Such work will have important bearing on the origins of radio continuum emission and the FIR-radio correlation.

\section{Acknowledgments}
 
 We are very grateful to Miller Goss for his support and insightful comments. We would also like to thank Amy Reines for useful discussions about radio flux measurements, Kelsey Johnson for providing some of the NGC 4449 radio data, and Deidre Hunter for supplying the H$\alpha$ images. This material is based upon work supported under a National Science Foundation Graduate Research Fellowship. We also acknowledge grant AST-0708002 from the National Science Foundation for funding the publication of this work. This research has made use of the NASA/IPAC Extragalactic Database (NED) which is operated by the Jet Propulsion Laboratory, California Institute of Technology, under contract with the National Aeronautics and Space Administration.

\newpage
\begin{figure}[htp]
\centering
\includegraphics[width=12cm]{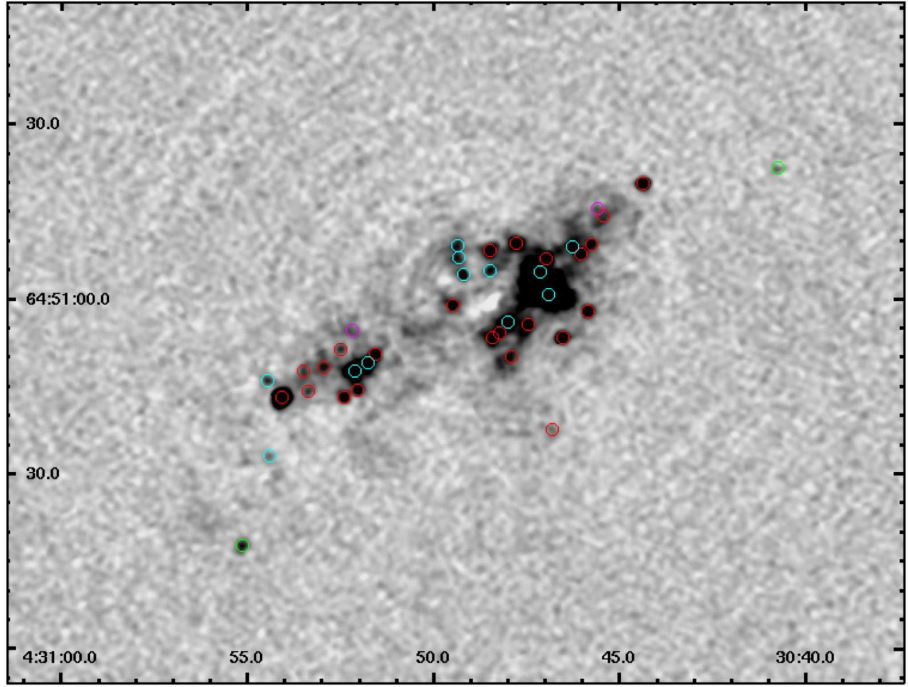}
\caption{20 cm image of the central region of NGC 1569, labeled with J2000 equatorial coordinates. Discrete 20 cm sources are marked with circles, and color-coded according to our classification. Green sources are background galaxies, red circles are SNR candidates, cyan circles are H~II regions, and magenta circles are sources which may be H~II regions or SNRs. Seven sources fall outside the bounds pictured here, all classified as background galaxies.}
\end{figure}

\begin{figure}[htp]
\centering
\includegraphics[width=12cm]{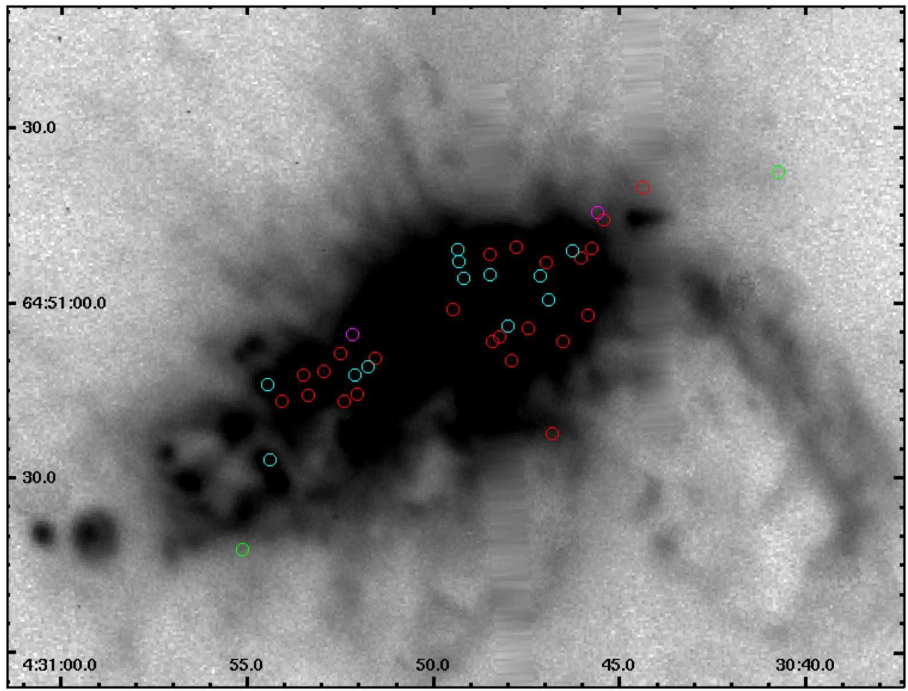}
\caption{The same region of NGC 1569 pictured in Figure 1, but in H$\alpha$ (image taken from Hunter \& Elmegreen 2004). Symbols are the same as in Figure 1.}
\end{figure}

\newpage
\begin{figure}[htp]
\centering
\includegraphics[width=12cm]{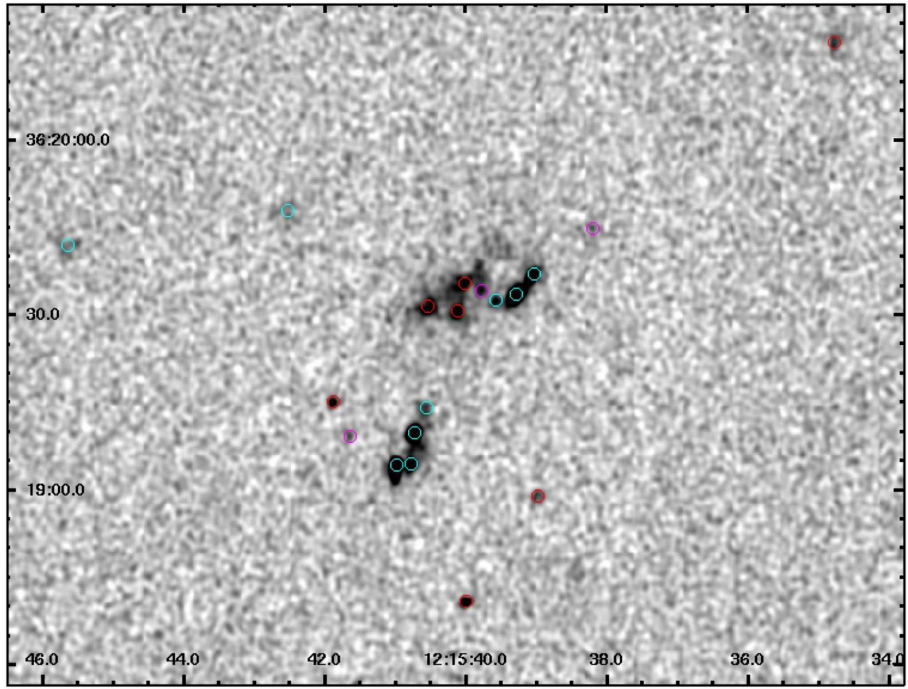}
\caption{20 cm image of the central region of NGC 4214. Colored circles have the same meaning as in Figure 1. Ten sources fall outside the bounds pictured here (nine background galaxies and one H~II region).}
\end{figure}

\begin{figure}[htp]
\centering
\includegraphics[width=12cm]{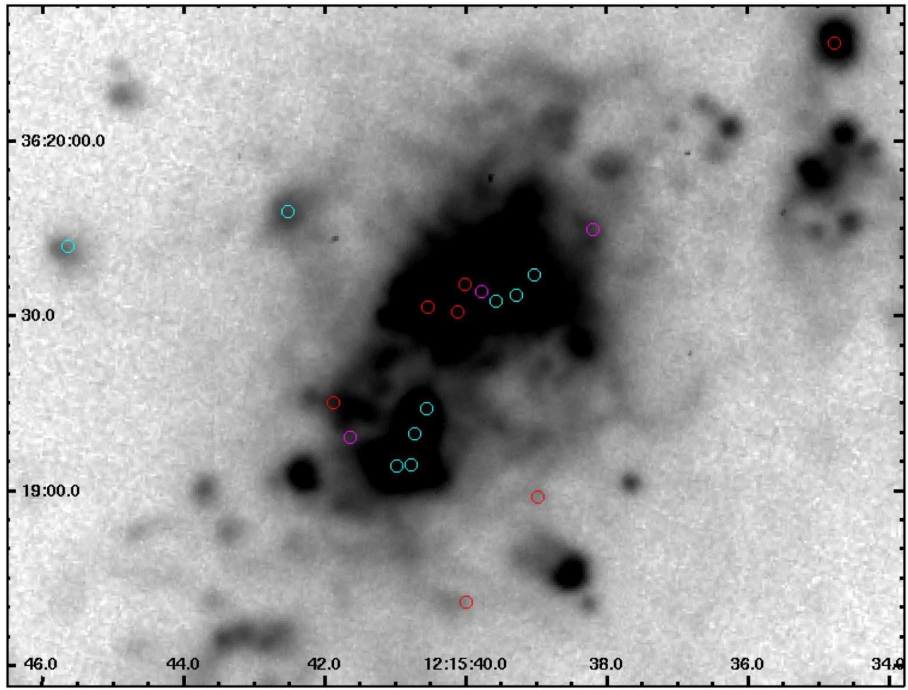}
\figcaption{H$\alpha$ image of NGC 4214 taken from Hunter \& Elmegreen (2004). Circles represent the same sources as in Figure 3, and are color-coded as described in Figure 1.}
\end{figure}

\newpage
\begin{figure}[htp]
\centering
\includegraphics[width=12cm]{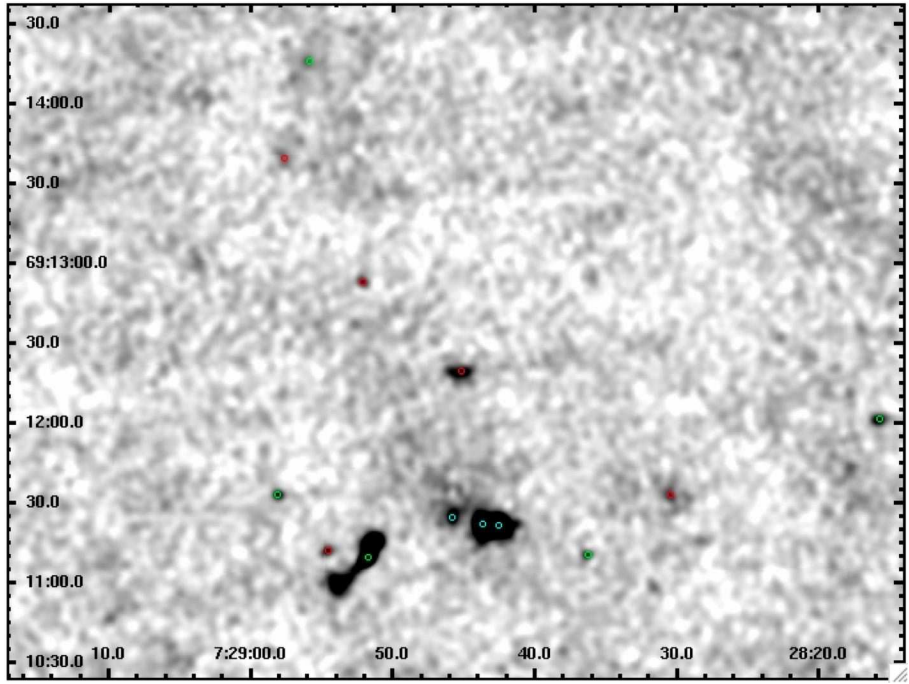}
\caption{20 cm image of NGC 2366, marked as described in Figure 1. Thirteen sources, all background galaxies, fall outside the bounds pictured here.}
\end{figure}

\begin{figure}[htp]
\centering
\includegraphics[width=12cm]{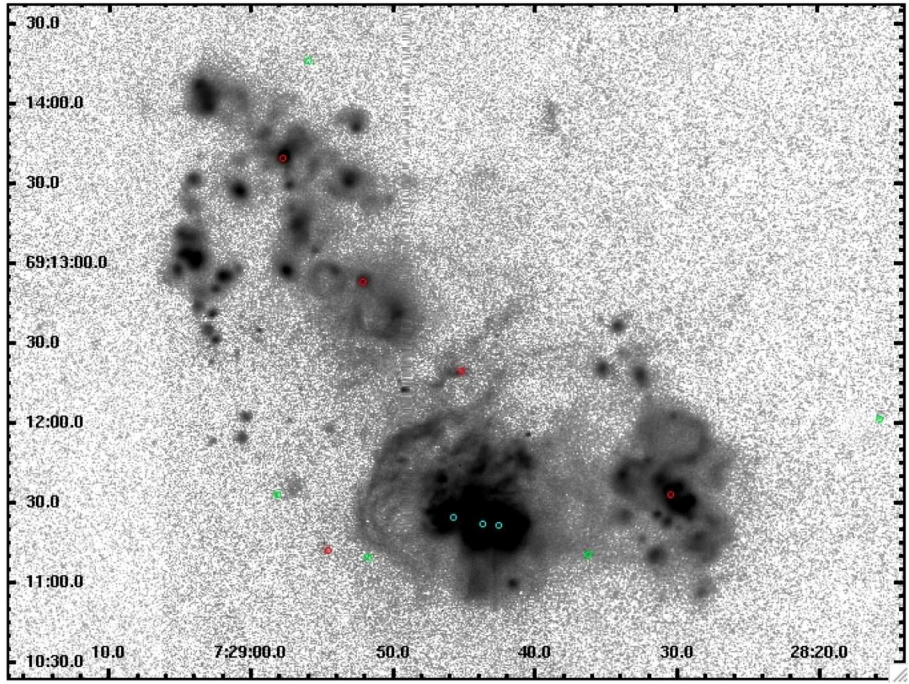}
\caption{H$\alpha$ image of NGC 2366 from Hunter \& Elmegreen (2004). Circles are the same as those in Figure 5, and their colors are described in Figure 1. }
\end{figure}

\newpage
\begin{figure}[htp]
\centering
\includegraphics[width=12cm]{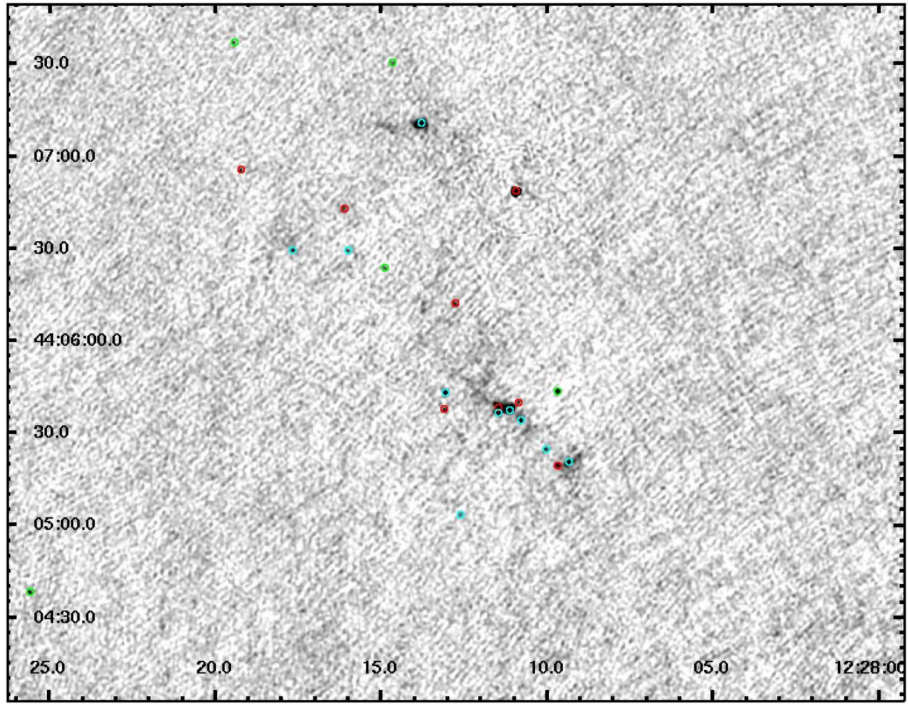}
\caption{20 cm image of NGC 4449, marked as described in Figure 1. Seven sources (classified here as background galaxies) fall outside the bounds pictured here.}
\end{figure}

\begin{figure}[htp]
\centering
\includegraphics[width=12cm]{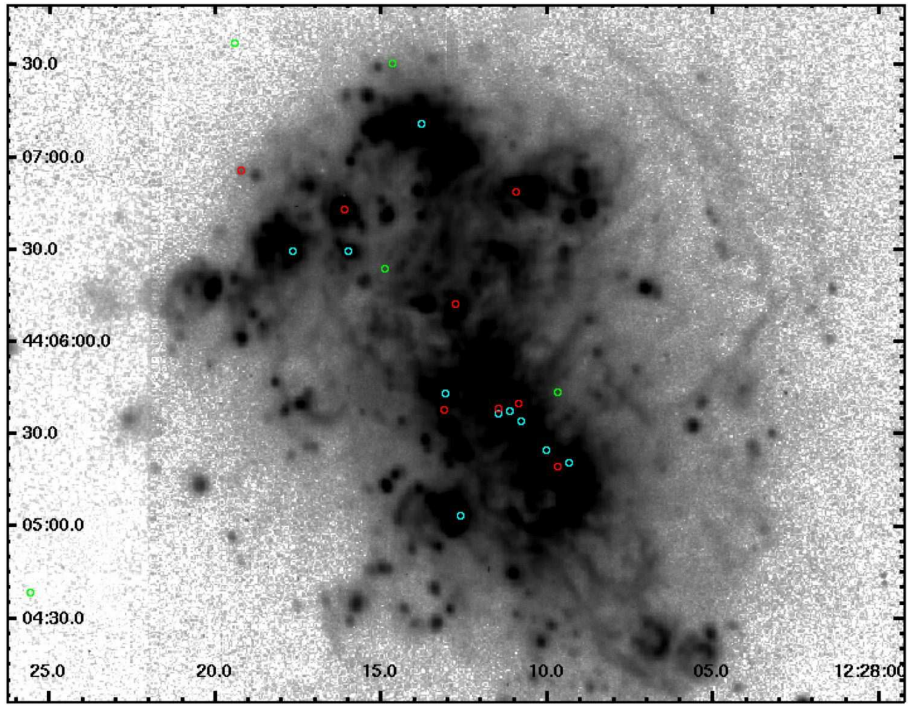}
\caption{H$\alpha$ image of NGC 4449 from Hunter \etal (1999). Circles represent the same sources as in Figure 7, and their colors are as described in Figure 1.}
\end{figure}

\newpage
\begin{figure}[htp]
\centering
\includegraphics[width=12cm]{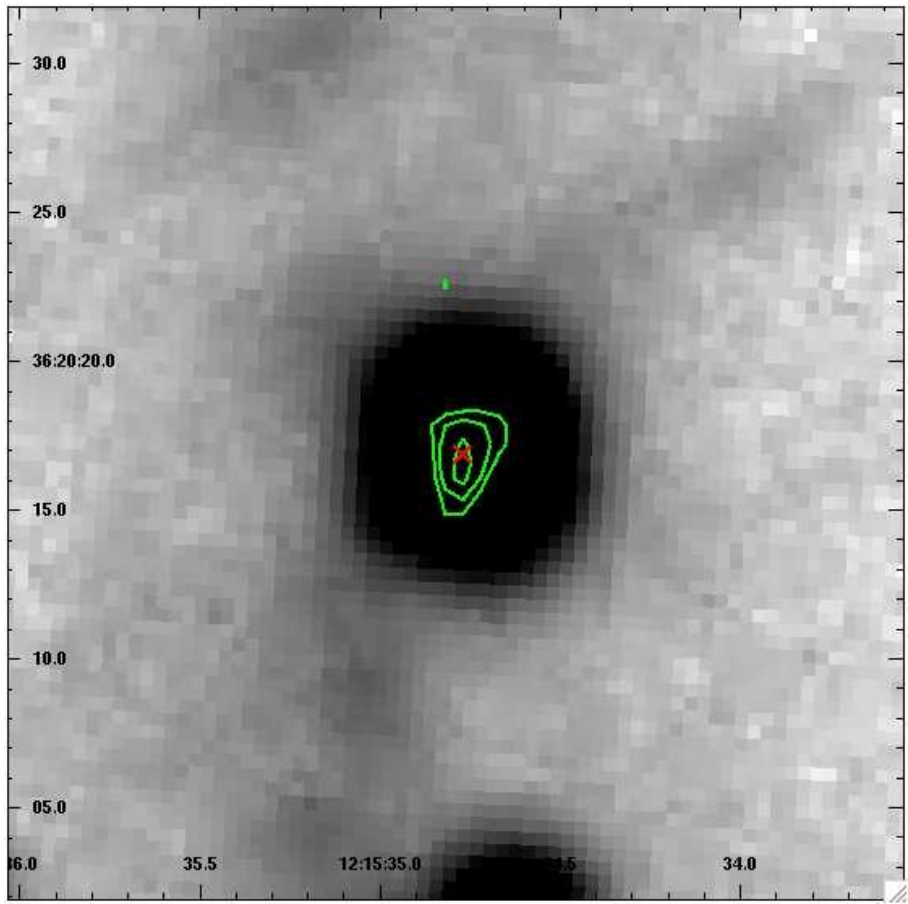}
\caption{Detail of NGC 4214 showing SNR candidate N4214-02. Grey-scale is an H$\alpha$ image taken from Hunter \& Elmegreen (2004) and green contours represent 20 cm radio continuum emission. The center of the SNR candidate is marked by a red X. The field-of-view is 30$^{\prime\prime}$ on a side. Contour levels are logarithmically spaced, representing 0.05, 0.11, and 0.18 mJy. Similar details of all radio SNR candidates will be available in the online version of the Astronomical Journal.}
\end{figure}

\newpage
\begin{deluxetable}{lccccccc}
\tablewidth{0 pt}
\tablecaption{ \label{tab:gals}
 Galaxy Parameters}
\tablehead{Galaxy & R.A. (2000)\tablenotemark{a}& Dec (2000)\tablenotemark{a} & Maj/Min Axis\tablenotemark{b} & Distance\tablenotemark{c} & M$_{B}$\tablenotemark{c} & S$_{20}$\tablenotemark{d} & log L$_{H\alpha}$\tablenotemark{e} \\
& (hr:min:sec) & ($^{\circ}$:\arcmin:\arcsec) & (arcmin) & (Mpc) & (mag) & (mJy) & (erg s$^{-1}$) }
\startdata
NGC 1569 & 04:30:49.0 & +64:50:53 & 3.3/2.0 & 2.0 & -17.1 & 411 & 40.5\\
NGC 4214 & 12:15:39.2 & +36:19:37 & 11.0/9.0 & 2.9 & -17.2 & 52 & 40.4 \\
NGC 2366 & 07:28:54.6 & +69:12:57 & 9.0/3.5 & 3.2 & -16.3 & 25 & 40.2 \\
NGC 4449 & 12:28:11.9 & +44:05:40 & 6.0/4.5 & 4.2 & -18.2 & 266 & 40.9\tablenotemark{f} \\
\enddata
\tablenotetext{a}{NED.}
\tablenotetext{b}{Nilson (1973).}
\tablenotetext{c}{\ Lee (2006).}
\tablenotetext{d}{Condon (1987).}
\tablenotetext{e}{Hunter \& Elmegreen (2004).}
\tablenotetext{f}{Hunter \etal (1999).}
\end{deluxetable} 
%
%
\begin{deluxetable}{lcccc}
\tablewidth{0 pt}
\tablecaption{
  VLA Data}
\tablehead{Galaxy & Band & Time On Source & Array Config. & Date Acquired \\
  & & (seconds) &  &  }
\startdata 
NGC 1569 & L & 14130 & A & 1983 Nov. 22 \\
& C & 10590 & B & 2001 May 13 \\
& X & 3910 & B & 1990 Aug 21 \\
& X & 10420 & C & 2001 July 26 \\
\hline
NGC 4214 & L & 35020 & A & 1986 May 3-4 \\
& C & 1460 & A & 1995 June 30 \\
& C &  860 & A & 2000 Oct. 20 \\
& C & 810 & BnA & 2001 Feb. 2 \\
& C & 1520 & B & 1997 April 20 \\
& X & 1510 & B & 1997 April 20\\
& X &   590 & C & 2001 July 9\\
\hline
NGC 2366 & L & 17750 & A & 1992 Nov. 23 \\
& L & 4700  & B & 1993 April 10 \\
& C & 54260 & C & 2006 Dec. 6-13 \\
& X & 11310 & B & 1993 April 10 \\
& X &  7110  & C & 1993 July 6 \\
\hline
NGC 4449 & L & 30340 & A & 1994 March 30 \\
& C &   2880 & A & 2002 Feb. 16 \\
& C &  1380 & B & 1984 Feb. 10 \\
& C &   2050 & B & 1985 May 2 \\
& C & 16080 & B(?) & 1994 June 16-17 \\ 
& X & 8550 & B & 2001 May 18 \\
& X & 13930 & C & 2001 June 20-Sep. 1 \\
\enddata
\end{deluxetable} 
\begin{deluxetable}{lcccc}
\tablewidth{0 pt}
\tablecaption{
Image Parameters}
\tablehead{Galaxy & Synthesized Beam & Spatial Res. & Freq. & Sensitivity \\
 & (arcsec) & (pc) & (GHz) & ($\mu$Jy/beam)}
\startdata 
NGC 1569 & 1.4 x 1.4 & 13 & 1.49 & 21 \\
 & & & 4.74 & 19 \\
 & & & 8.21 & 19 \\
\hline
NGC 4214 & 1.35 x 1.35 &19 & 1.49 & 19 \\
 & & & 4.86 & 41 \\
 & & & 8.46 & 38 \\
\hline
NGC 2366 & 3.7 x 3.7 & 57 & 1.43 & 22 \\
 & & & 4.86 & 20 \\
 & & & 8.44 & 16 \\
\hline
NGC 4449 & 1.35 x 1.35 & 28 & 1.45 & 25 \\
 & & & 4.86 & 24 \\
 & & & 8.46 & 15 \\
\enddata
\end{deluxetable} 
%
%
\begin{deluxetable}{lccccccccc}
\tablewidth{0 pt}
\tablecaption{ 
20 cm Sources in NGC 1569}
\tablehead{ID & RA & Dec & Maj/Min Axis & S$_{20}$ & S$_{6}$ & S$_{3.6}$ & $\alpha$ & H$\alpha$? & Class \\
 & (hr:min:sec) & ($^{\circ}$:\arcmin:\arcsec) & (arcsec) & (mJy) & (mJy) & (mJy) & & & }
\startdata 
N1569-01 & 4:30:01.50 & 64:52:53.3  &   7.4/1.6   & $0.63\pm0.14$ & $<0.36$ &   --- & $<-0.48$ & --- & Bkg ($>$R$_{25}$) \\
N1569-02 & 4:30:26.32 &  64:48:33.4  &   5.2/1.4  & $0.42\pm0.09$ & $<0.25$ & $0.53\pm0.13$ & $0.14\pm0.19$ & No & Bkg ($>$R$_{25}$) \\
N1569-03 & 4:30:40.75 &  64:51:22.6  &   2.0/1.4  & $0.18\pm0.03$ & $0.12\pm0.03$ & $<0.37$ & $-0.36\pm0.28$ & No & Bkg \\
N1569-04 & 4:30:44.35 &  64:51:20.1  &   1.7/1.5  & $1.17\pm0.05$ & $0.62\pm0.04$ & $0.47\pm0.06$ & $-0.54\pm0.10$ & Yes & SNR \\

N1569-05 & 4:30:45.32 &  64:51:14.9  &   4.9/2.0  & $0.70\pm0.14$ & $0.40\pm0.09$ & $<0.51$ & $-0.48\pm0.26$ & Yes & SNR \\
N1569-06 & 4:30:45.60 &  64:51:15.8  &   1.5/1.4  & $0.09\pm0.03$ & $<0.09$ & $<0.12$ & $<0.00$ & Yes & SNR/H~II \\
N1569-07 & 4:30:45.75 &  64:51:09.6  &   1.6/1.4  & $0.23\pm0.07$ & $0.10\pm0.03$ & $<0.15$ & $-0.74\pm0.39$ & Yes & SNR \\ 
N1569-08 & 4:30:45.79 &  64:50:58.2  &   1.7/1.6  & $0.75\pm0.08$ & $0.36\pm0.03$ & $0.24\pm0.04$ & $-0.68\pm0.12$ & Yes & SNR \\

N1569-09 & 4:30:46.03 &  64:51:08.0  &   1.6/1.5  & $0.30\pm0.08$ & $0.22\pm0.03$ & $0.12\pm0.04$ & $-0.60\pm0.11$ & Yes & SNR \\
N1569-10 & 4:30:46.28 &  64:51:09.0  &   1.5/1.4  & $0.19\pm0.04$ & $0.17\pm0.03$ & $0.14\pm0.03$ & $-0.17\pm0.07$ & Yes & H~II\tablenotemark{a} \\
N1569-11 & 4:30:46.54 &  64:50:53.6  &   1.6/1.4  & $1.01\pm0.07$ & $0.48\pm0.06$ & $0.41\pm0.07$ & $-0.55\pm0.13$ & Yes & SNR \\
N1569-12 & 4:30:46.83 &  64:50:37.8  &   2.3/2.0  & $0.28\pm0.08$ & $<0.19$ & $<0.23$ & $<-0.34$ & Yes & SNR \\

N1569-13 & 4:30:46.93 &  64:51:00.8  &   5.3/2.2  & $10.44\pm0.32$ & $9.20\pm0.25$ & $9.19\pm0.25$ & $-0.08\pm0.54$ & Yes & H~II \\
N1569-14 & 4:30:46.97 &  64:51:07.2  &   1.5/1.4  & $1.30\pm0.09$ & $0.57\pm0.07$ & $0.36\pm0.04$ & $-0.76\pm0.14$ &  Yes & SNR\tablenotemark{b} \\
N1569-15 & 4:30:47.15 &  64:51:04.7  &   4.2/2.9  & $6.31\pm0.32$ & $5.70\pm0.32$ & $5.11\pm0.31$ & $-0.12\pm0.59$ & Yes & H~II \\
N1569-16 & 4:30:47.48 &  64:50:55.7  &   2.1/1.7  & $0.50\pm0.11$ & $0.37\pm0.08$ & $0.32\pm0.08$ & $-0.27\pm0.19$ & Yes & SNR \\

N1569-17 & 4:30:47.73 &  64:51:09.7  &   2.3/1.8  & $0.88\pm0.16$ & $0.67\pm0.13$ & $0.53\pm0.15$ & $-0.29\pm0.29$ & Yes & SNR \\
N1569-18 & 4:30:47.91 &  64:50:50.3  &   2.9/1.5  & $0.37\pm0.12$ & $<0.37$ & $0.27\pm0.08$ & $-0.20\pm0.25$ & Yes & SNR \\
N1569-19 & 4:30:47.99 &  64:50:56.5  &   2.0/1.8  & $1.11\pm0.11$ & $0.96\pm0.13$ & $0.79\pm0.11$ & $-0.19\pm0.21$ & Yes & H~II\tablenotemark{a} \\
N1569-20 & 4:30:48.20 &  64:50:54.7  &   1.5/1.3  & $0.27\pm0.08$ & $0.22\pm0.05$ & $0.16\pm0.04$ & $-0.31\pm0.12$ & Yes & SNR \\

N1569-21 & 4:30:48.42 &  64:50:53.6  &   2.7/2.3  & $0.82\pm0.16$ & $0.67\pm0.13$ & $0.52\pm0.12$ & $-0.26\pm0.27$ & Yes & SNR \\
N1569-22 & 4:30:48.44 &  64:51:05.3  &   1.4/1.4  & $0.23\pm0.06$ & $0.26\pm0.05$ & $0.27\pm0.05$ & $0.11\pm0.10$ & Yes & H~II \\
N1569-23 & 4:30:48.48 &  64:51:08.5  &   1.8/1.6  & $0.47\pm0.12$ & $0.33\pm0.10$ & $0.33\pm0.10$ & $-0.21\pm0.21$ & Yes & SNR \\
N1569-24 & 4:30:49.19 &  64:51:04.6  &   1.5/1.3  & $0.44\pm0.08$ & $0.54\pm0.08$ & $0.43\pm0.08$ & $0.02\pm0.15$ & Yes & H~II \\

N1569-25 & 4:30:49.35 &  64:51:07.2  &   2.0/2.0  & $0.42\pm0.09$ & $0.45\pm0.08$ & $0.41\pm0.09$ & $0.01\pm0.16$ & Yes & H~II \\
N1569-26 & 4:30:49.36 &  64:51:09.2  &   2.0/2.0  & $0.39\pm0.09$ & $0.36\pm0.08$ & $0.32\pm0.09$ & $-0.11\pm0.16$ & Yes & H~II \\
N1569-27 & 4:30:49.50 &  64:50:59.4  &   3.2/2.8  & $1.46\pm0.23$ & $0.95\pm0.16$ & $0.92\pm0.17$ & $-0.28\pm0.39$ & Yes & SNR\tablenotemark{b} \\
N1569-28 & 4:30:51.55 &  64:50:51.0  &   1.7/1.4  & $0.56\pm0.08$ & $0.29\pm0.04$ & $0.18\pm0.04$ & $-0.67\pm0.12$ & Yes & SNR\tablenotemark{b} \\

N1569-29 & 4:30:51.77 &  64:50:49.2  &   3.1/2.5  & $1.40\pm0.17$ & $1.04\pm0.12$ & $1.02\pm0.12$ & $-0.19\pm0.28$ & Yes & H~II \\
N1569-30 & 4:30:52.04 &  64:50:44.4  &   1.8/1.4  & $0.40\pm0.10$ & $0.23\pm0.05$ & $0.25\pm0.05$ & $-0.23\pm0.15$ & Yes & SNR \\
N1569-31 & 4:30:52.15 &  64:50:47.8  &   3.2/3.2  & $2.66\pm0.18$ & $2.30\pm0.15$ & $2.24\pm0.15$ & $-0.10\pm0.32$ & Yes & H~II \\
N1569-32 & 4:30:52.19 &  64:50:54.8  &   3.4/2.2  & $0.57\pm0.17$ & $<0.51$ & $<0.52$ & $<-0.11$ & Yes & SNR/H~II \\

N1569-33 & 4:30:52.42 &  64:50:43.2  &   1.7/1.5  & $0.49\pm0.10$ & $0.25\pm0.08$ & $<0.33$ & $-0.59\pm0.31$ & Yes & SNR \\
N1569-34 & 4:30:52.46 &  64:50:51.6  &   1.9/1.4  & $0.29\pm0.08$ & $<0.17$ & $<0.17$ & $<-0.44$ & Yes & SNR\\
N1569-35 & 4:30:52.96 &  64:50:48.8  &   2.1/1.5  & $0.25\pm0.08$ & $0.23\pm0.04$ & $0.17\pm0.04$ & $-0.25\pm0.11$ & Yes & SNR \\
N1569-36 & 4:30:53.36 &  64:50:44.5  &   1.7/1.4  & $0.22\pm0.04$ & $0.15\pm0.03$ & $<0.11$ & $-0.29\pm0.20$ & Yes & SNR \\

N1569-37 & 4:30:53.53 &  64:50:47.7  &   3.2/1.6  & $0.32\pm0.10$ & $0.25\pm0.05$ & $0.16\pm0.05$ & $-0.46\pm0.15$ & Yes & SNR \\
N1569-38 & 4:30:54.08 &  64:50:43.5  &   2.1/1.9  & $4.67\pm0.08$ & $2.48\pm0.07$ & $1.71\pm0.08$ & $-0.58\pm0.15$ & Yes & SNR\tablenotemark{b} \\
N1569-39 & 4:30:54.38 &  64:50:33.5  &   2.1/1.5  & $0.19\pm0.06$ & $0.17\pm0.05$ & $0.16\pm0.03$ & $-0.12\pm0.09$ & Yes & H~II \\
N1569-40 & 4:30:54.43 &  64:50:46.3  &   1.8/1.3  & $0.26\pm0.06$ & $0.22\pm0.05$ & $0.26\pm0.05$ & $-0.02\pm0.10$ & Yes & H~II \\

N1569-41 & 4:30:55.14 &  64:50:17.8  &   2.1/1.7  & $0.51\pm0.06$ & $0.30\pm0.05$ & $<0.19$ & $-0.46\pm0.18$ & No & Bkg \\
N1569-42 & 4:30:58.00 &  64:49:12.2  &   3.1/1.4  & $0.94\pm0.06$ & $1.39\pm0.04$ & $1.31\pm0.06$ & $0.23\pm0.11$ & No & Bkg ($>$R$_{25}$) \\
N1569-43 & 4:31:03.74 &  64:54:28.6  &   4.8/1.3  & $0.37\pm0.08$ & $0.85\pm0.07$ &  --- & $0.73\pm0.19$ & --- & Bkg ($>$R$_{25}$) \\
N1569-44 & 4:31:04.40 &  64:55:36.7  &   7.1/1.4  & $0.91\pm0.12$ & $0.39\pm0.10$ &  --- & $-0.74\pm0.25$ & --- & Bkg ($>$R$_{25}$) \\

N1569-45 & 4:31:13.04 &  64:52:29.3  &   3.9/1.4  & $1.03\pm0.07$ & $1.08\pm0.05$ & $0.61\pm0.09$ & $-0.17\pm0.14$ & No & Bkg ($>$R$_{25}$) \\
N1569-46 & 4:31:16.05 &  64:49:09.4  &   5.3/1.6  & $1.63\pm0.10$ & $0.88\pm0.05$ & $0.62\pm0.14$ & $-0.55\pm0.20$ & --- & Bkg ($>$R$_{25}$) \\
\enddata
\tablenotetext{a}{This source is actually likely an SNR based on [Fe~II] and Pa$\beta$ emission in Labrie \& Pritchet (2006).}
\tablenotetext{b}{Additional data in the literature lend credence to this characterization; See Section 6.}
\end{deluxetable} 
%
%
\begin{deluxetable}{lccccccccc}
\tablewidth{0 pt}
\tablecaption{ 
20 cm Sources in NGC 4214}
\tablehead{ID & R.A. (2000) & Dec. (2000) & Maj/Min Axis & S$_{20}$ & S$_{6}$ & S$_{3.6}$ & $\alpha$ & H$\alpha$? & Class \\
 & (hr:min:sec) & ($^{\circ}$:\arcmin:\arcsec) & (arcsec) & (mJy) & (mJy) & (mJy) & & & }
\startdata 
N4214-01 & 12:15:28.66 & 36:23:02.2 &  7.4/1.5  & $0.86\pm0.13$ & $<0.45$ &  --- & $<-0.54$ & No & Bkg \\
N4214-02 & 12:15:34.74 & 36:20:17.1 &  3.6/2.3  & $0.47\pm0.10$ & $0.34\pm0.06$ & $0.21\pm0.05$ & $-0.48\pm0.14$ & Yes & SNR\\
N4214-03 & 12:15:38.18 & 36:19:44.9 &  3.5/1.4  & $0.18\pm0.05$ & $<0.21$ & $<0.18$ & $<0.00$ & Yes & SNR/H~II \\
N4214-04 & 12:15:38.98 & 36:18:59.1 &  2.4/1.8  & $0.28\pm0.06$ & $0.16\pm0.05$ & $<0.18$ & $-0.48\pm0.31$ & Yes & SNR \\

N4214-05 & 12:15:39.04 & 36:19:37.1 &  2.7/1.9  & $0.88\pm0.07$ & $0.88\pm0.12$ & $0.77\pm0.08$ & $-0.07\pm0.14$ & Yes & H~II \\
N4214-06 & 12:15:39.29 & 36:19:33.7 &  3.2/2.0  & $1.80\pm0.08$ & $1.98\pm0.16$ & $1.61\pm0.10$ & $-0.05\pm0.17$ & Yes & H~II \\
N4214-07 & 12:15:39.55 & 36:19:32.7 &  1.7/1.3  & $0.16\pm0.05$ & $<0.16$ & $0.26\pm0.04$ & $0.29\pm0.19$ & Yes & H~II \\
N4214-08 & 12:15:39.78 & 36:19:34.3 &  3.0/1.4  & $0.16\pm0.04$ & $<0.22$ & $<0.14$ & $<-0.08$ & Yes & SNR/H~II \\

N4214-09 & 12:15:39.99 & 36:18:41.1 &  1.8/1.4  & $0.46\pm0.04$ & $0.28\pm0.04$ & $0.16\pm0.03$ & $-0.62\pm0.07$ & Yes & SNR \\
N4214-10 &  12:15:40.02 & 36:19:35.5 &  6.0/3.4 & $1.43\pm0.22$ & $<1.38$ & $<0.75$ & $<-0.37$ & Yes & SNR \\
N4214-11 &  12:15:40.12 & 36:19:30.7 &  3.1/2.4 & $0.66\pm0.10$ & $<0.74$ & $0.26\pm0.065$ & $-0.53\pm0.17$ & Yes & SNR \\
N4214-12 &  12:15:40.55 & 36:19:31.5 &  4.3/3.1 & $1.06\pm0.16$ & $0.74\pm0.21$ & $0.40\pm0.12$ & $-0.56\pm0.26$ & Yes & SNR \\

N4214-13 & 12:15:40.56 & 36:19:14.1 &  1.4/1.4  & $0.13\pm0.02$ & $0.11\pm0.03$ & $0.16\pm0.03$ & $0.03\pm0.04$ & Yes & H~II \\
N4214-14 & 12:15:40.73 & 36:19:09.9 & 2.1/1.6 & $1.44\pm0.05$ & $1.44\pm0.09$ & $1.30\pm0.07$ & $-0.05\pm0.11$ & Yes & H~II \\
N4214-15 & 12:15:40.76 & 36:19:04.5 & 1.9/1.4  & $0.10\pm0.03$ & $0.19\pm0.03$ & $0.10\pm0.03$ & $0.09\pm0.05$ & Yes & H~II \\
N4214-16 &  12:15:40.98 & 36:19:04.3 & 2.5/1.8 & $1.74\pm0.06$ & $1.79\pm0.11$ & $1.58\pm0.08$ & $-0.04\pm0.13$ & Yes & H~II \\

N4214-17 & 12:15:41.36 & 36:21:14.1 &  6.4/1.3  & $0.33\pm0.09$ & $<0.34$ & $0.27\pm0.06$ & $-0.11\pm0.20$ & Yes & H~II \\
N4214-18 & 12:15:41.64 & 36:19:09.7 &  2.1/1.5  & $0.15\pm0.05$ & $<0.19$ & $<0.15$ & $<0.00$ & Yes & SNR/H~II \\
N4214-19 & 12:15:41.87 & 36:19:15.3 &  1.8/1.4  & $0.30\pm0.04$ & $<0.37$ & $<0.14$ & $<-0.43$ & Yes & SNR \\
N4214-20 & 12:15:42.52 & 36:19:47.9 &  4.4/3.8  & $0.53\pm0.17$ & $0.59\pm0.15$ & $<0.48$ & $0.10\pm0.35$ & Yes & H~II \\

N4214-21 & 12:15:44.65 & 36:21:20.1 &  3.0/1.3  & $0.16\pm0.05$ & $<0.13$ & $<0.12$ & $<-0.16$ & No & Bkg \\
N4214-22 & 12:15:44.98 & 36:17:56.3 &  1.9/1.4  & $0.12\pm0.03$ & $0.20\pm0.06$ & $<0.17$ & $0.46\pm0.31$ & No & Bkg \\
N4214-23 & 12:15:45.64 & 36:19:42.1 &  3.9/3.1  & $0.43\pm0.12$ & $0.48\pm0.10$ & $<0.73$ & $0.10\pm0.30$ & Yes & H~II \\
N4214-24 & 12:15:46.87 & 36:18:55.5 &  1.9/1.4  & $0.17\pm0.04$ & $0.39\pm0.09$ & $0.20\pm0.04$ & $0.14\pm0.08$ & No & Bkg \\

N4214-25 & 12:15:48.56 & 36:21:55.7 &  5.8/3.3  & $1.16\pm0.20$ & $1.11\pm0.17$ & $0.70\pm0.21$ & $-0.24\pm0.37$ & No & Bkg \\
N4214-26 & 12:15:49.24 & 36:21:45.8 &  5.2/1.7  & $0.44\pm0.11$ & $<0.34$ & $<0.41$ & $<-0.24$ & No & Bkg \\
N4214-27 & 12:15:49.68 & 36:18:47.7 &  2.4/1.4  & $0.53\pm0.05$ & $<0.45$ & $0.16\pm0.04$ & $-0.69\pm0.14$ & No & Bkg \\
N4214-28 & 12:15:51.15 & 36:18:45.0 &  2.3/1.2  & $0.24\pm0.04$ & $<0.67$ & $<0.14$ & $<-0.32$ & No & Bkg \\
N4214-29 & 12:15:57.18 & 36:18:30.0 &  4.6/1.3  & $1.15\pm0.08$ & $0.53\pm0.07$ & --- & $-0.66\pm0.13$ & No & Bkg \\
\enddata
\end{deluxetable} 
%
%
\begin{deluxetable}{lccccccccc}
\tablewidth{0 pt}
\tablecaption{ 20 cm Sources in NGC 2366}
\tablehead{ID & R.A. (2000) & Dec. (2000) & Maj/Min Axis & S$_{20}$ & S$_{6}$ & S$_{3.6}$ & $\alpha$ & H$\alpha$? & Class \\
 & (hr:min:sec) & ($^{\circ}$:\arcmin:\arcsec) & (arcsec) & (mJy) & (mJy) & (mJy) & & & }
\startdata 
N2366-01 & 7:27:48.04 & 69:14:48.7  &   9.7/3.7   & $ 2.96\pm0.07 $ & $ 1.30\pm0.13$ &  --- & $-0.67\pm0.08$ & No & Bkg ($>$R$_{25}$) \\
N2366-02 & 7:28:00.71 & 69:08:40.5  &   6.9/4.0   & $ 0.28\pm0.07 $ & $<0.42$ & $<0.57$ & $<0.34$ & --- & Bkg ($>$R$_{25}$) \\   
N2366-03 & 7:28:05.57 & 69:12:35.3  &   4.9/4.2   & $ 0.29\pm0.05 $ & $0.56\pm0.06$ & $0.56\pm0.07$ & $0.41\pm0.10$ & No & Bkg ($>$R$_{25}$) \\   
N2366-04 & 7:28:06.29 & 69:08:19.7  &  11.1/5.7  & $ 0.57\pm0.13 $ & $ <0.86$ & $<1.07$ & $<0.34$ & --- & Bkg ($>$R$_{25}$) \\

N2366-05 & 7:28:15.73 & 69:12:01.7  &   5.0/3.8   & $ 0.33\pm0.05 $ & $0.12\pm0.03$ & $<0.13$ & $-0.83\pm0.23$ & No & Bkg ($>$R$_{25}$) \\
N2366-06 & 7:28:24.16 & 69:18:21.7  &  17.6/4.5  & $ 0.60\pm0.13 $ & $<0.57$ & --- & $<-0.05$ & --- & Bkg ($>$R$_{25}$) \\
N2366-07 & 7:28:30.41 & 69:11:33.8  &   4.3/3.7   & $ 0.20\pm0.02 $ & $0.10\pm0.03$ & $0.08\pm0.02$ & $-0.53\pm0.04$ & Yes & SNR \\
N2366-08 & 7:28:34.59 & 69:17:24.6  &  10.6/6.1  & $ 0.90\pm0.13 $ & $0.29\pm0.06$ & --- & $-0.92\pm0.21$ & --- & Bkg ($>$R$_{25}$) \\

N2366-09 & 7:28:36.27 & 69:11:10.9  &   4.0/3.3   & $ 0.16\pm0.04 $ & $<0.10$ & $<0.07$ & $<-0.48$ & No & Bkg \\
N2366-10 & 7:28:42.58 & 69:11:22.0  &   5.0/4.6   & $ 6.50\pm0.06 $ & $6.27\pm0.05$ & $5.16\pm0.04$ & $-0.13\pm0.09$ & Yes & H~II \\
N2366-11 & 7:28:43.71 & 69:11:22.4  &   7.4/4.6   & $ 2.75\pm0.08 $ & $2.51\pm0.07$ & $2.32\pm0.05$ & $-0.10\pm0.12$ & Yes & H~II \\
N2366-12 & 7:28:45.26 & 69:12:19.8  &   9.1/5.9   & $ 0.92\pm0.12 $ & $0.30\pm0.07$ & $<0.27$ & $-0.91\pm0.21$ & Yes & SNR \\

N2366-13 & 7:28:45.69 & 69:11:25.8  &   3.8/3.7   & $ 0.23\pm0.03 $ & $0.23\pm0.02$ & $0.19\pm0.02$ & $-0.12\pm0.04$ & Yes & H~II \\
N2366-14 & 7:28:51.75 & 69:11:09.9  & 30.0/10.0 & $ 5.19\pm0.18 $ & $2.10\pm0.11$ & $1.49\pm0.10$ & $-0.70\pm0.27$ & No & Bkg \\
N2366-15 & 7:28:52.10 & 69:12:54.4  &   4.1/3.7   & $ 0.19\pm0.05 $ & $0.06\pm0.02$ & $0.08\pm0.02$ & $-0.37\pm0.07$ & Yes & SNR \\
N2366-16 & 7:28:54.57 & 69:11:12.7  &   4.3/3.7   & $ 0.15\pm0.03 $ & $0.13\pm0.03$ & $0.09\pm0.02$ & $-0.29\pm0.04$ & Yes & SNR \\

N2366-17 & 7:28:55.98 & 69:14:16.6  &   7.0/3.7   & $ 0.21\pm0.05 $ & $<0.09$ & $<0.19$ & $<-0.67$ & No & Bkg \\
N2366-18 & 7:28:57.67 & 69:13:41.0  &   5.5/3.7   & $ 0.19\pm0.04 $ & $0.11\pm0.02  $ & $<0.13$ & $-0.42\pm0.21$ & Yes & SNR \\
N2366-19 & 7:28:58.00 & 69:11:33.9  &   4.6/4.1   & $ 0.28\pm0.05 $ & $<0.13$ & $<0.09$ & $<-0.63$ & No & Bkg \\
N2366-20 & 7:29:13.46 & 69:10:16.7  &   4.9/3.9   & $ 0.24\pm0.05 $ & $<0.14$ & $<0.17$ & $<-0.44$ & No & Bkg ($>$R$_{25}$) \\

N2366-21 & 7:29:17.10 & 69:09:05.9  &   7.9/4.0   & $ 0.28\pm0.07 $ & $<0.19$ & $<0.34$ & $<-0.34$ & No & Bkg ($>$R$_{25}$) \\
N2366-22 & 7:29:26.78 & 69:12:28.2  &   6.4/4.1   & $ 0.75\pm0.06 $ & $ 0.18\pm0.04  $ & $<0.38$ & $-1.19\pm0.20$ & No & Bkg ($>$R$_{25}$) \\
N2366-23 & 7:29:28.80 & 69:16:56.0  &  11.8/4.0  & $ 0.71\pm0.11 $ & $<0.38$ & --- & $<-0.52$ & No & Bkg ($>$R$_{25}$) \\
N2366-24 & 7:29:32.60 & 69:11:53.7  &   8.1/3.3   & $ 0.29\pm0.07 $ & $<0.16$ & $<0.49$ & $<-0.49$ & No & Bkg ($>$R$_{25}$) \\ 
\enddata
\end{deluxetable} 
%
%
\begin{deluxetable}{lccccccccc}
\tablewidth{0 pt}
\tablecaption{ 
20 cm Sources in NGC 4449}
\tablehead{ID & R.A. (2000) & Dec. (2000) & Maj/Min Axis & S$_{20}$ & S$_{6}$ & S$_{3.6}$ & $\alpha$ & H$\alpha$? & Class \\
 & (hr:min:sec) & ($^{\circ}$:\arcmin:\arcsec) & (arcsec) & (mJy) & (mJy) & (mJy) & & & }
\startdata 
N4449-01 & 12:27:56.35 &  44:03:06.1 & 1.5/1.2  & $0.20\pm0.06$ & $0.18\pm0.05$ & $<0.29$ & $-0.10\pm0.32$ & --- & Bkg ($>$R$_{25}$) \\
N4449-02 & 12:28:00.39 &  44:07:52.9 & 1.6/1.4  & $0.21\pm0.05$ & $<0.23 $ & $ 0.25\pm0.03$ & $0.09\pm0.14$ & No & Bkg ($>$R$_{25}$) \\
N4449-03 & 12:28:06.96 &  44:08:18.2 & 1.8/1.8  & $0.37\pm0.04$ & $0.22\pm0.04$ & $0.17\pm0.03$ & $-0.44\pm0.07$ & --- & Bkg ($>$R$_{25}$) \\
N4449-04 & 12:28:06.98 &  44:08:20.0 & 1.4/1.4  & $0.22\pm0.03$ & $<0.10$ & $<0.08$ & $<-0.67$ & --- & Bkg ($>$R$_{25}$) \\

N4449-05 & 12:28:07.05 &  44:08:16.8 & 2.7/1.3  & $0.23\pm0.07$ & $<0.19$ & $<0.15$ & $<-0.24$ & --- & Bkg ($>$R$_{25}$) \\
N4449-06 & 12:28:09.37 &  44:05:20.6 & 3.1/2.0  & $0.68\pm0.09$ & $0.69\pm0.10$ & $0.57\pm0.05$ & $-0.10\pm0.13$ & Yes & H~II \\
N4449-07 & 12:28:09.67 &  44:05:19.8 & 1.4/1.4  & $0.24\pm0.03$ & $0.16\pm0.03$ & $0.10\pm0.02$ & $-0.51\pm0.04$ & Yes & SNR \\
N4449-08 & 12:28:09.69 &  44:05:43.8 & 1.4/1.3  & $0.75\pm0.05$ & $0.40\pm0.04$ & $0.27\pm0.03$ & $-0.59\pm0.07$ & No & Bkg \\

N4449-09 & 12:28:10.04 &  44:05:25.0 & 1.8/1.3  & $0.19\pm0.06$ & $0.18\pm0.03$ & $0.14\pm0.04$ & $-0.17\pm0.09$ & Yes & H~II \\
N4449-10 & 12:28:10.76 &  44:05:34.2 & 1.5/1.1  & $0.17\pm0.05$ & $0.29\pm0.06$ & $0.25\pm0.04$ & $0.23\pm0.08$ & Yes & H~II \\
N4449-11 & 12:28:10.87 &  44:05:40.2 & 1.4/1.4  & $0.13\pm0.03$ & $<0.11$ & $<0.07$ & $<-0.35$ & Yes & SNR \\
N4449-12\tablenotemark{a} & 12:28:10.93 &  44:06:48.8 & 1.4/1.3  & $9.66\pm0.05$ & $4.84\pm0.04$ & $2.11\pm0.02$ & $-0.90\pm0.07$ & Yes & SNR\tablenotemark{b} \\

N4449-13 & 12:28:11.12 &  44:05:38.0 & 2.5/2.3  & $1.56\pm0.12$ & $1.46\pm0.12$ & $1.47\pm0.09$ & $-0.03\pm0.19$ & Yes & H~II \\
N4449-14 & 12:28:11.47 &  44:05:38.6 & 3.2/2.7  & $1.46\pm0.16$ & $1.00\pm0.08$ & $0.64\pm0.04$ & $-0.54\pm0.19$ & Yes & SNR \\
N4449-15 & 12:28:11.49 &  44:05:37.0 & 2.2/1.1  & $0.30\pm0.06$ & $0.24\pm0.05$ & $0.21\pm0.04$ & $-0.19\pm0.09$ & Yes & H~II \\
N4449-16 & 12:28:12.63 &  44:05:03.6 & 2.0/1.1  & $0.15\pm0.05$ & $0.17\pm0.03$ & $0.13\pm0.02$ & $-0.12\pm0.06$ & Yes & H~II \\

N4449-17 & 12:28:12.77 &  44:06:12.2 & 1.8/1.6  & $0.25\pm0.06$ & $0.14\pm0.04$ & $0.07\pm0.02$ & $-0.78\pm0.08$ & Yes & SNR \\
N4449-18 & 12:28:13.05 &  44:05:43.4 & 1.9/1.5  & $0.46\pm0.06$ & $0.62\pm0.07$ & $0.55\pm0.04$ & $0.11\pm0.10$ & Yes & H~II \\
N4449-19 & 12:28:13.07 &  44:05:37.8 & 1.6/1.6  & $0.23\pm0.06$ & $<0.15$ & $<0.09$ & $<-0.55$ & Yes & SNR \\
N4449-20 & 12:28:13.82 &  44:07:10.8 & 3.1/2.3  & $1.89\pm0.12$ & $1.64\pm0.09$ & $1.80\pm0.08$ & $-0.03\pm0.18$ & Yes & H~II \\

N4449-21 & 12:28:14.66 &  44:07:30.6 & 1.6/1.4  & $0.14\pm0.04$ & $<0.11$ & $<0.08$ & $<-0.28$ & No & Bkg \\
N4449-22 & 12:28:14.91 &  44:06:24.0 & 1.9/0.9  & $0.14\pm0.04$ & $<0.10$ & $0.08\pm0.01$ & $-0.33\pm0.19$ & No & Bkg \\
N4449-23 & 12:28:15.98 &  44:06:29.5 & 1.9/1.3  & $0.20\pm0.05$ & $0.26\pm0.05$ & $0.20\pm0.02$ & $-0.04\pm0.07$ & Yes & H~II \\
N4449-24 & 12:28:16.13 &  44:06:43.3 & 2.6/2.3  & $0.37\pm0.10$ & $0.44\pm0.07$ & $0.19\pm0.04$ & $-0.53\pm0.13$ & Yes & SNR \\

N4449-25 & 12:28:17.67 &  44:06:29.7 & 2.8/1.9  & $0.37\pm0.09$ & $0.39\pm0.10$ & $0.34\pm0.04$ & $-0.05\pm0.13$ & Yes & H~II \\
N4449-26 & 12:28:19.23 &  44:06:55.9 & 1.4/1.4  & $0.16\pm0.02$ & $<0.10$ & $0.10\pm0.02$ & $-0.26\pm0.11$ & Yes & SNR \\
N4449-27 & 12:28:19.45 &  44:07:37.1 & 1.5/1.2  & $0.14\pm0.04$ & $0.22\pm0.04$ & $0.19\pm0.03$ & $0.19\pm0.07$ & No & Bkg \\
N4449-28 & 12:28:23.69 &  44:03:13.7 & 2.7/1.3  & $0.20\pm0.06$ & $0.26\pm0.07$ & $<0.35$ & $0.24\pm0.35$ & No & Bkg ($>$R$_{25}$) \\

N4449-29 & 12:28:25.58 &  44:04:38.5 & 1.4/1.3  & $0.21\pm0.04$ & $0.10\pm0.03$ & $<0.12$ & $-0.67\pm0.31$ & No & Bkg ($>$R$_{25}$) \\
N4449-30 & 12:28:25.90 &  44:03:19.1 & 1.8/1.3  & $0.29\pm0.05$ & $0.25\pm0.05$ & $0.31\pm0.07$ & $-0.03\pm0.10$ & No & Bkg ($>$R$_{25}$) \\
\enddata
\tablenotetext{a}{The flux of this source is time variable, and therefore, given the different epochs of the flux measurements, the spectral index is not trustworthy.}
\tablenotetext{b}{Additional data in the literature lend credence to this characterization; See Section 6.}
\end{deluxetable} 
%
%
\begin{deluxetable}{lccccc}                                            
\tablewidth{0 pt}
\tablecaption{The Components of Radio Continuum Emission}
\tablehead{Galaxy & S$_{20}$(H~II$_{20})$\tablenotemark{a} & S$_{20}$(H~II$_{H\alpha}$)\tablenotemark{b} & S$_{20}$(th)\tablenotemark{c} & S$_{20}$(SNR)\tablenotemark{d} & S$_{20}$(nt)\tablenotemark{e} \\
  & (mJy) & (mJy) & (mJy) & (mJy) & (mJy) }
\startdata 
NGC 1569  & $>$24.0 & $>$53.5 & 95 & $>$17.7 & 316\\
NGC 2366  & $>$9.5 & $>$8.3 & $<$25 & $>$1.7 & $<$25 \\
NGC 4449  & $>$6.0 & $>$13.7& 61 & $>$12.5 & 205\\
\enddata
\tablenotetext{a}{Total 20 cm flux density for the H~II regions identified in this work.}
\tablenotetext{b}{Based on the H$\alpha$-selected catalogs of H~II regions of Kennicutt \etal (1989) and Buckalew \& Kobulnicky (2006), H$\alpha$ fluxes were summed and then converted to thermal 20 cm radio emission.}
\tablenotetext{c}{Integrated thermal 20 cm emission, calculated using the total 20 cm flux densities in Condon (1987) and thermal fractions from Niklas \etal (1997) and Lisenfeld \etal (2004).}
\tablenotetext{d}{Total 20 cm flux density for the SNRs identified in this work.}
\tablenotetext{e}{Integrated non-thermal 20 cm emission, calculated using the same resources as column \emph{c}.}
\end{deluxetable}

\end{document}